\documentclass[pre,twocolumn]{revtex4}
\usepackage{amsmath,amssymb,latexsym,epsfig,graphics,epsf}
\usepackage{graphics}
\usepackage{float}

\newcommand{\be}{\begin{equation}}
\newcommand{\ee}{\end{equation}}

\begin{document}
\title{Isomorph invariance of Couette shear flows simulated by the SLLOD equations of motion}
\author{Leila Separdar}
\affiliation{Department of Physics, College of Sciences, Shiraz University, Shiraz 71454, Iran}
\author{Nicholas P. Bailey}
\affiliation{DNRF Center "Glass and Time", IMFUFA, Dept. of Sciences, Roskilde University, P. O. Box 260, DK-4000 Roskilde, Denmark}
\author{Thomas B. Schr{\o}der}
\affiliation{DNRF Center "Glass and Time", IMFUFA, Dept. of Sciences, Roskilde University, P. O. Box 260, DK-4000 Roskilde, Denmark}
\author{Saeid Davatolhagh}
\affiliation{Department of Physics, College of Sciences, Shiraz University, Shiraz 71454, Iran}
\author{Jeppe C. Dyre}\email{dyre@ruc.dk}
\affiliation{DNRF Center "Glass and Time", IMFUFA, Dept. of Sciences, Roskilde University, P. O. Box 260, DK-4000 Roskilde, Denmark}

\date{\today}

\begin{abstract}
Non-equilibrium molecular dynamics simulations were performed to study the thermodynamic, structural, and dynamical properties of the single-component Lennard-Jones and the Kob-Andersen binary Lennard-Jones liquids. Both systems are known to have strong correlations between equilibrium thermal fluctuations of virial and potential energy. Such systems have good isomorphs (curves in the thermodynamic phase diagram along which structural, dynamical, and some thermodynamic quantities are invariant when expressed in reduced units). The SLLOD equations of motion were used to simulate Couette shear flows of the two systems. We show analytically that these equations are isomorph invariant provided the reduced strain rate is fixed along the isomorph. Since isomorph invariance is generally only approximate, a range of shear rates were simulated to test for the predicted invariance, covering both the linear and non-linear regimes. For both systems, when represented in reduced units the radial distribution function and the intermediate scattering function are identical  for state points that are isomorphic. The strain-rate dependence of the viscosity, which exhibits shear thinning, is also invariant along an isomorph. Our results extend the isomorph concept to the non-equilibrium situation of a shear flow, in which the phase diagram is three dimensional because the shear rate defines a third dimension.
\end{abstract}
\maketitle

\section{ Introduction}\label{intro}

Investigating liquids in non-equilibrium situations has been a matter of interest both theoretically and numerically in recent decades. Statistical mechanics provides a link between microscopic states and equilibrium thermodynamics, but in non-equilibrium situations it is difficult to find such a link\cite{Evans/Morriss:2008}. General formalisms for nonlinear response theory include the transient time-correlation formalism\cite{Morriss/Evans:1987} and the Kawasaki formalism\cite{Yamada/Kawasaki:1975}. Many theories for describing non-equilibrium liquids have been motivated by the theories of equilibrium situations and of the glassy states. Fluctuation-dissipation relation violations\cite{fdr}, mode-coupling theory\cite{MCT}, and dynamical heterogeneity\cite{hetero}, are examples of frameworks used to describe the behavior of systems under homogeneous shear flow. 

According to the isomorph theory\cite{I,II,III,IV,V}, the class of so-called strongly correlating liquids have isomorphs, which are curves in the phase diagram along which structural, dynamical, and some thermodynamic properties are invariant when expressed in reduced units. This theory has been tested successfully experimentally\cite{nature} and numerically\cite{pedersen,gna,pedersen2}. In all cases studied so far in detail the systems were in equilibrium, however. The above-mentioned interest in non-equilibrium situations motivated us to investigate whether the isomorph theory -- or an extension thereof -- holds in situations involving non-equilibrium steady states. To address this question we performed non-equilibrium molecular dynamics simulations on two different systems, the single component Lennard-Jones (SCLJ) liquid and the Kob-Andersen binary Lennard-Jones (KABLJ) liquid\cite{Kob/Andersen:1994, Kob/Andersen:1995a,Kob/Andersen:1995b}. Shear flow has been previously investigated in both systems; the SCLJ system was studied in Refs.~\onlinecite{Ge,shearSCLJ,travisSCLJ} and the KABLJ system in Refs.~\onlinecite{fdr,butler,hoover}. In the NEMD simulations reported below we focus on homogeneous flows generated by the SLLOD equations of motion proposed by Evans, Morriss, and Ladd some time ago\cite{moris,ladd,Evans/Morriss:2008}. 

The paper is organized as follows. In Sec. \ref{theor_back} we briefly describe the theoretical basis of this work, specifically the SLLOD equations, and the isomorph theory. Section \ref{isom_inv_SLLOD} proves that the SLLOD equations of motion are isomorph invariant if the isomorph concept is extended to a three-dimensional phase diagram in which the shear rate defines the third dimension. In Sec. \ref{generating_isom} the procedure for generating isomorphic steady state points is explained. Models and simulation details are presented in Sec. \ref{model_sim}. The results of the simulations are presented in Sec. \ref{sim_results}. The paper concludes with a brief discussion in Sec. \ref{summary}.

\section{ Theoretical background}\label{theor_back}

\subsection{The SLLOD equations of motion}

Non-equilibrium molecular dynamics (NEMD) techniques have been used extensively to study homogeneous and inhomogeneous fluids under the influence of different flow fields. The case of a homogeneous shear flow was among the first applications of NEMD\cite{hoover/ashurst}; it was later generalized to elongational flows\cite{elong}. Two issues arise when simulating shear flows. The first is that any algorithm must ensure that the shear viscosity $\eta\equiv\sigma_{xy}/\dot{\gamma}$ at low strain rates $\dot\gamma$ obeys the Green-Kubo linear-response relation -- here $V$ is the volume, $T$ the temperature, and $\sigma_{xy}$ is the $xy$ element of the spatially averaged stress tensor, i.e., $\sigma_{xy}\equiv \sum_i x_i F_{y,i}/V$ where $x_i$ is the x-coordinate of the i'th particle and $F_{y,i}$ is the y-coordinate of the force on this particle\cite{hoover},

\be 
\eta = \frac{V}{k_BT}\int^{\infty}_{0} \langle\sigma_{xy}(0)\sigma_{xy}(t)\rangle dt\,.
\ee
The second issue arising when simulation a shear flow is that flow generates heat. In order to simulate a steady viscous flow this heat must be removed, which is typically done using a thermostat. For homogeneous NEMD a commonly used method is the so-called Gaussian\cite{gausian} thermostat based on time-reversible constraint forces, which keeps either the total energy (ergostat) or the kinetic energy (isokinetic thermostat) fixed. Another popular method is the Nos\'{e}-Hoover thermostat, which uses an additional dynamical variable to simulate the heat bath.

Two well-known algorithms for simulating viscous shear flow are the DOLLS and SLLOD algorithms. The first homogeneous NEMD algorithm was based on the DOLLS Hamiltonian proposed by Hoover {\it et al.}\cite{doll},

\be 
H_{\rm DOLLS}
\,=\,U(\mathbf{r}_1, ...,\mathbf{r}_N )+\sum_i \mathbf{p}^{2}_{i}/2m_i+\sum_i \mathbf{r}_{i}\cdot\nabla\mathbf{v}\cdot\mathbf{p}_{i}\, .
\ee
Here $U$ is the potential energy of the system consisting of $N$ particles, $\nabla\mathbf{v}$ is the gradient tensor of the macroscopic streaming velocity field $\mathbf{v}(\mathbf{r})$, $m_i$ is the mass of particle $i$, $\mathbf{r}_i$ and $\mathbf{p_i}$ are, respectively, its laboratory position and ``peculiar'' (thermal) momentum. The latter quantity relates to the velocity $\mathbf{c_i}$ relative to the streaming velocity field $\mathbf{v}(\mathbf{r})$ via $\mathbf{p}_i\equiv m_i\mathbf{c}_i$, where $\mathbf{c}_i$ is given by

\be 
\mathbf{c}_i = \mathbf{v}_i-\mathbf{v(r}_i)\,.
\ee
Via the standard Hamilton equations of motion, the equations generated from the DOLLS Hamiltonian are 

\begin{eqnarray} 
\label{dolls1} 
\mathbf{\dot{r}}_i&=&\mathbf{p}_i/m_i+\mathbf{r}_i\cdot\nabla\mathbf{v} \\
\mathbf{\dot{p}}_i&=&\mathbf{F}_i-\nabla\mathbf{v}\cdot\mathbf{p}_i\label{dolls2}\,.
\end{eqnarray} 
Here $\mathbf{F}_i$ is the force exerted on each particle by the surrounding particles. It was shown, however, by Evans and Morriss\cite{Evans/Morriss:2008} that these equations are only suitable for simulating flows in the linear-response regime. Evans and Morriss\cite{moris} and Ladd\cite{ladd} have shown that more suitable equations for generating flows in both the linear and non-linear regimes are 

\begin{eqnarray} 
\label{sllod1}
\mathbf{\dot{r}}_i&=&\mathbf{p}_i/m_i+\mathbf{r}_i\cdot\nabla\mathbf{v} \\
\mathbf{\dot{p}}_i&=&\mathbf{F}_i-\mathbf{p}_i\cdot\nabla\mathbf{v}\label{sllod2}\,. 
\end{eqnarray} 
The macroscopic streaming velocity field is assumed to have a linear profile, i.e., a constant spatial gradient. The difference between these equations and the DOLLS equations lies in the second term in Eqs.\ \eqref{dolls2} and \eqref{sllod2}, which has been transposed -- thus the name was also ``transposed'' from DOLLS to SLLOD. These equations of motion plus the Lees-Edwards boundary conditions\cite{lees}, in conjunction with a Gaussian kinetic thermostat, guarantee that homogeneous flows in both the linear and non-linear regimes are generated, although it has been shown recently that the flow generated by SLLOD still exhibits certain differences compared to the more physical boundary-driven flow\cite{hoover}. Excellent reviews of methods for simulating homogeneous flows can be found in Refs. \onlinecite{hoover} and \onlinecite{todSLLOD}. The SLLOD equations of motion were recently used by Edan and Procaccia to study zero-temperature plastic flows of amorphous solids\cite{ler09}.

A special case of the SLLOD equations of motion is the Couette shear flow, where all elements of the strain-rate tensor are zero except one,

\be
\nabla \mathbf{v}= \begin{pmatrix}
\frac{\partial v_x}{\partial x} & \frac{\partial v_y}{\partial x} & \frac{\partial v_z}{\partial x}\\
\frac{\partial v_x}{\partial y} & \frac{\partial v_y}{\partial y} & \frac{\partial v_z}{\partial y}\,.\\
\frac{\partial v_x}{\partial z} & \frac{\partial v_y}{\partial z} & \frac{\partial v_z}{\partial z}
\end{pmatrix}
=
\begin{pmatrix}
0 & 0 & 0\\
\dot{\gamma} & 0 & 0\\
0 & 0 & 0
\end{pmatrix}\,.
\ee
Here $\dot\gamma$ is the shear rate, i.e., the gradient in the $y$-direction of the $x$-component of the streaming velocity field. Substituting the above strain rate tensor into Eqs.\ \eqref{sllod1} and \eqref{sllod2}, they take the following simple forms:
\begin{eqnarray} 
\dot{\mathbf{r}}_i&=& \mathbf{p}_i/m_i+ \mathbf{i}\mathbf{\dot{\gamma}} y_i \\
\mathbf{\dot{p}}_i&=& {\mathbf{F}_i}-\mathbf{i}\dot{\gamma}p_{yi}\,. 
\end{eqnarray}
Here ${\bf i}$ is the unit vector in the positive $x$ axis direction.

\subsection{The isomorph theory and its predictions}\label{isom_theory}

From the instantaneous positions and momenta of all particles one can find the instantaneous total energy and pressure from

\begin{eqnarray} 
E&=&K(\mathbf{p}_1, \ldots, \mathbf{p}_N) + U(\mathbf{r}_1, \ldots, \mathbf{r}_N) \\
pV&=&Nk_BT(\mathbf{p}_1, \ldots, \mathbf{p}_N)+W(\mathbf{r}_1, \ldots, \mathbf{r}_N)\,. 
\end{eqnarray} 
Here $K(\mathbf{p}_1, \ldots, \mathbf{p}_N)$ and $U(\mathbf{r}_1, \ldots, \mathbf{r}_N)$ are the kinetic and potential energies, respectively, $T(\mathbf{p}_1, \ldots, \mathbf{p}_N)$ is the instantaneous kinetic temperature given by the kinetic energy per particle, and $W(\mathbf{r}_1, \ldots, \mathbf{r}_N)\equiv-1/3\sum_i\mathbf{r}_i.\nabla_{\mathbf{r}_i}U(\mathbf{r}_1, \ldots, \mathbf{r}_N)$ is the instantaneous virial, which after dividing by volume is the configurational contribution to the instantaneous pressure.

From the fluctuations of potential energy and virial two parameters can be defined\cite{I,II,III,IV,V}: the density-scaling exponent $\gamma$ (this name is explained in Sec. \ref{generating_isom}),

\be\label{gamma1}
\gamma\equiv \frac{\langle\Delta W\Delta U\rangle}{\langle(\Delta U)^{2}\rangle},
\ee
and the correlation coefficient $R$, 

\be 
R=\frac{\langle\Delta W \Delta U\rangle}{\sqrt{\langle(\Delta W)^2\rangle\langle(\Delta U)^2\rangle}}. 
\ee
In both expressions the angular brackets denote an NVT ensemble average referring to a given thermodynamic state point. Liquids that have $R\geq0.9$ \cite{I,II} are simple in the Roskilde meaning of the term\cite{prx}. For Roskilde simple liquids it is possible to find (approximate) isomorphs in the thermodynamic phase diagram, defined as follows. Consider two state points (1) and (2) with densities $\rho_1$ and $\rho_2 $ and temperatures $T_1$ and $T_2$, respectively. These state points are by definition isomorphic\cite{IV} if any two of their respective microscopic configurations, whose coordinates scale into each other according to

\be 
\rho_1^{1/3} \mathbf{r}_i^{(1)} = \rho_2^{1/3} \mathbf{r}_i^{(2)}\,\,\,(i=1,...,N)\,,
\ee
to a good approximation have proportional Boltzmann statistical weights:

\be 
e^{-U(\mathbf{r}_1^{(1)},\ldots,\mathbf{r}_N^{(1)})/k_BT_1}= C_{12} e^{-U(\mathbf{r}_i^{(2)},\ldots,\mathbf{r}_N^{(2)})/k_BT_2}\,. 
\ee
Here $C_{12}$ is a proportionality constant that depends only on the two state points, not on the microscopic configurations. Pairs of state points in the phase diagram that are isomorphic fall onto the same isomorphic curve, for brevity termed an  ``isomorph''; an isomorph is thus an equivalence class of state points. While the existence of isomorphs is typically only approximate (except for inverse-power-law systems), the theory has been developed as a set of consequences of the above definition; these can then be systematically investigated in simulations and experiments.

In reduced units, as a result of the proportionality of Boltzmann factors, state points on an isomorph have the same dynamic, structural, and (some) thermodynamic quantities\cite{IV}. Reduced units refer to the state point by giving lengths in units of $\rho^{-1/3}$, time in units of $\rho^{-1/3}(k_BT/m)^{-1/2}$, and energy in units of $k_BT$. The invariant thermodynamic  quantities include\cite{IV} the excess entropy (the difference between the entropy of the liquid and of the corresponding ideal gas at same density and temperature) and the isochoric specific heat. The structure is invariant along an isomorph in reduced coordinates. The equilibrium dynamic properties are also invariant; normalized time-autocorrelation functions, average relaxation times $\tau_A\equiv{\int}^\infty\langle A(0)A(t)\rangle dt/\langle A^2\rangle$,  and the intermediate scattering function are examples of dynamical quantities invariant along an isomorph when expressed in reduced units\cite{IV}.

The use of reduced units may remind the reader of the principle of corresponding states. It is textbook knowledge that the two parameters in the classical van der Waals equation of state allow identification of different substances' phase diagrams by considering scaled versions of the temperature and pressure (or temperature and density). The same applies for the two parameters in the Lennard-Jones potential; in fact in simulations one typically uses LJ units (also called MD units) where energies including $k_BT$ and lengths are scaled by the parameters $\epsilon$ and $\sigma$, respectively -- then there is only one single-component Lennard-Jones potential. It is also possible to relate the properties of mixtures to those of a single-component fluid via appropriate mixing rules, which determine effective values of energy and length parameters\cite{Henderson/Leonard:1971,Hanley/Evans:1981}. We emphasize, however, that such scaling arguments say nothing about the existence of isomorphs in the phase diagram of a given system, which effectively reduces the equilibrium phase diagram to one dimension for all quantities that are isomorph invariant. On the other hand, Rosenfeld discussed excess entropy scaling as a kind of corresponding states principle\cite{Rosenfeld:1999}; the isomorph theory is similar in spirit to this idea, because for a given system a generalized excess entropy scaling follows from the existence of isomorphs\cite{IV}. Rosenfeld's original motivation was variational hard-sphere perturbation theory, in which the effective hard-sphere  diameter is the only relevant variable, making also the phase diagram effectively one-dimensional.

A good starting point for determining whether a quantity is isomorph invariant is given by Eq.\ (9) of Ref. \onlinecite{IV} valid for the most important microconfigurations (the tilde signals reduced coordinates -- $\tilde {\bf r}_i\equiv \rho^{1/3}{\bf r}_i$ -- and $Q$ labels the state point)

\be \label{isom}
U({\bf r}_{1}, \ldots, {\bf r}_N; \rho) = k_BT f_1(\tilde {\bf r}_1,\ldots, \tilde {\bf r}_N) + g(Q)\,.
\ee
This follows directly from the isomorph definition. When we generalize the isomorph concept below to non-equilibrium situations, the term $g(Q)$ may also depend on the strain rate since this quantity defines a third state-point coordinate.

\subsection{Isomorph invariance of the SLLOD equations of motion }\label{isom_inv_SLLOD}

The SLLOD equations of motion are given by Eqs.\ \eqref{sllod1} and \eqref{sllod2}. In order to show that these equations are isomorph invariant, one needs to substitute all quantities in terms of their reduced forms -- the thermodynamically scaled dimensionless forms, denoted by tilde. Most fundamental are the scaling of lengths by $\rho^{-1/3}$ and energies by $k_B T$: thus $\mathbf{r}_i=\tilde{\mathbf{r}}_i \rho^{-1/3}$ and $U = \tilde U k_B T$, whereas the scaling of time depends on the dynamics. If inertia is present, as in ordinary Newtonian dynamics and in the SLLOD equations, we scale the particle masses by their average value $m$, i.e., $m_i=\tilde{m}_i m$. The scaling of time then follows from requiring invariance of the zero-strain rate equations of motion \cite{IV}. To see this we make the above replacements in the SLLOD equations, denoting the time scaling factor to be determined as $t_0$ ($t=\tilde{t}t_0$). The scaling factor of momentum is $m\rho^{-1/3}/t_0$ (we omit the isokinetic thermostatting term here for simplicity; it can also be properly expressed in reduced form). From the first SLLOD equation we have (in which $\dot{\tilde{\mathbf{r}}}_i=d\tilde{\mathbf{r}}_i/d\tilde{t}$)

\begin{equation} 
  \dot{\tilde{\mathbf{r}}}_i \frac{\rho^{-1/3}}{t_0} = 
  \frac{\tilde{\mathbf{p}}_i}{\tilde m_i} \frac{m\rho^{-1/3}}{t_0 m} + 
  \tilde{\mathbf{r}}_i\cdot \widetilde{\nabla} \tilde{\mathbf{v}}  \frac{\rho^{-1/3}}{t_0}\,.
\end{equation}
Clearly, all scaling factors cancel whatever choice is made for $t_0$, implying isomorph invariance independent of $t_0$. From the second SLLOD equation we have

\begin{equation}
\dot{\tilde{\mathbf{p}}}_i \frac{m\rho^{-1/3}}{{t_0}^2} = 
- \tilde\nabla_i \tilde U k_B T \rho^{1/3} - 
\tilde{\mathbf{p}}_i \cdot \widetilde{\nabla} \tilde{\mathbf{v}} \frac{m\rho^{-1/3}}{{t_0}^2},
\end{equation}
where $\widetilde\nabla_i$ denotes the gradient with respect to $\tilde{\mathbf{r}}_i$. Here, in order to be able to cancel the scaling factors, it is necessary that $ k_B T \rho^{1/3} = m\rho^{-1/3}/{t_0}^2$, or

\begin{equation}
t_0 =  \sqrt{m /k_B T} \rho^{-1/3},
\end{equation}
as also shown to be the case for Newtonian dynamics in Ref.~\onlinecite{IV}. The reduced SLLOD equations are thus:

\be
\label{S1} \dot{\tilde{\mathbf{r}}}_i=\tilde{\mathbf{p}}_i/\tilde{m}_i+\tilde{\mathbf{r}}_i\cdot\widetilde{\nabla}\tilde{\mathbf{v}}, 
\ee
\be
\label{S2} \dot{\tilde{\mathbf{p}}}_i=\tilde{\mathbf{F}}_i-\mathbf{\tilde{p}}_i\cdot\widetilde{\nabla}\tilde{\mathbf{v}}\,. 
\ee
The isomorph invariance now follows from the fact that along an isomorph $\tilde{\mathbf{F}}_{i}$ is a unique function of reduced coordinates (which follows from Eq. (\ref{isom}) of Ref. \onlinecite{IV}); this applies when the reduced strain rate tensor $\widetilde{\nabla} \tilde{\mathbf{v}}=\nabla {\mathbf v}\, t_0$ is fixed. This means that along an isomorph the strain rate must vary in such a way that its reduced form is fixed, see also Eq.\ \eqref{gamma} below. In the three-dimensional phase diagram parameterized by density, temperature, and strain rate, the isomorphs are one-dimensional curves. Thus it takes two parameters to specify an isomorph, for example the excess entropy and the reduced strain rate, in contrast to standard equilibrium thermodynamic isomorphs that are labelled by just one parameter \cite{IV}.

For any given isomorph in the $(\rho,T,\dot\gamma)$ phase diagram one can also consider its projection onto the $(\rho,T)$ plane, which we call the ``projected  isomorph''. {\it A priori}, one cannot expect this projection to coincide with an equilibrium isomorph. In fact, one can imagine starting at a given $(\rho,T)$ point with many different strain rates, and tracing out different isomorphs. Their projections will be in general different, except in the limit of low strain rate.  Empirically, though, we do find that the projected isomorphs coincide with each other and with the equilibrium isomorph containing the original $(\rho,T)$ point.

\subsection{Generating isomorphic state points}\label{generating_isom}

It was shown recently\cite{trond} that for simple liquids and solids, temperature can be written as a product of a function of the excess entropy per particle, $s$, and a function of density: $T=f(s)h(\rho)$. Accordingly, one can generate curves of constant excess entropy by requiring\cite{trond,boh12} that 

\be\label{3} 
\frac{h(\rho)}{T}={\rm Const}.
\ee 
It is only in Roskilde simple (i.e., strongly correlating) liquids that these configurational adiabats are also isomorphs, i. e., have the property that all the other isomorph invariants apply. 

The function $h(\rho)$ is called the density-scaling function; its logarithmic derivative is the density-scaling exponent $\gamma$\cite{pedersen,IV}, which is also given by Eq. (\ref{gamma1}),

\be\label{gama} 
\gamma\equiv \left(\frac{\partial \ln T}{\partial \ln \rho}\right)_{s}=\,\frac{d\ln h}{d\ln \rho}\,.
\ee
It follows that $\gamma$ only depends on the density. Note that in this paper $\gamma$ is always the density-scaling exponent, whereas $\dot\gamma$ is always the strain rate. 

Another consequence of the isomorph theory is an expression for $h(\rho)$ for atomic liquids with interaction potentials consisting of a sum of inverse power laws, $v(r)=\sum_n v_n r^{-n}$. For such liquids $h(\rho)$ is given as follows\cite{trond,boh12}

\be\label{heq}
h(\rho)=\sum_n C_n \rho^{n/3}\,,
\ee
where the only non-zero terms are those corresponding to an $r^{-n}$ term in the pair potential. For Lennard-Jones liquids the pair potential has the form

\be\label{LJ} 
v(r)=4\epsilon\left[\left(\frac{\sigma}{r}\right)^{12}-\left(\frac{\sigma}{r}\right)^6\right]\,.
\ee
It follows from Eq. \eqref{heq} that the density scaling function can be written as $h(\rho)=A\rho^4-B\rho^2$; thus LJ isomorphs are given by an expression of the form

\be\label{S3} 
\frac{A\rho^4-B\rho^2}{T}={\rm Const}\,. 
\ee

Since $h(\rho)$ is defined up to a multiplicative factor, we are free to choose a particular normalization yielding a one-parameter expression. Starting from a reference state point $(\rho_0$,$T_0$,$\dot{\gamma}_0)$, plotting the instantaneous virial versus the instantaneous potential energy determines via a least-squares linear fit the value of $\gamma$ according to Eq.\ \eqref{gamma1}, denoted by $\gamma_0$. Following Refs.~\onlinecite{trond} and \onlinecite{boh12} we can write $h(\rho)=\alpha\tilde\rho^4+(1-\alpha)\tilde\rho^2$ where $\tilde\rho\equiv\rho/\rho_0$; since $\gamma=d\ln h/d\ln\rho$ it follows that $\gamma_0=2\alpha+2$, i.e., $\alpha=\gamma_0/2-1$. Consequently

\be\label{H} 
h=(\gamma_0/2-1)\tilde\rho^4-(\gamma_0/2-2)\tilde\rho^2 \,.
\ee
To generate the isomorph of the reference state point we repeatedly used the equation (in which $\tilde T\equiv T/T_0$) 

\be\label{T}  
\tilde T= h(\tilde\rho)\,,
\ee 
keeping the reduced strain rate fixed via (where $\dot\gamma_0$ is the strain rate at the reference state point)

\be\label{gamma} 
\dot{\gamma}=\dot{\gamma}_0\tilde\rho^{1/3}\tilde T^{1/2}\,.
\ee

As mentioned above, isomorph invariance is only approximate. Moreover, one cannot be certain that the above expression for $h(\rho)$ applies in non-equilibrium situations -- the parameters could depend on $\dot\gamma_0$.

\section{Model and details of simulation}\label{model_sim}

To test the invariance of the SLLOD equations in practice two standard simple liquids were simulated for a range of shear rates, covering both the linear and non-linear regimes. The two atomic systems SCLJ (single-component Lennard-Jones) and KABLJ (Kob-Andersen binary Lennard-Jones mixture) were simulated. In the SCLJ system $500$ particles interact via the Lennard-Jones (LJ) potential Eq. \eqref{LJ}. In the unit system where $\sigma=1$ and $\epsilon=1$ (so-called LJ units) reference state points for generating isomorphs are given by $\rho_0=0.84$, $T_0=0.8$, for several strain rates up to $2.5$. The potential was shifted and truncated at $3.5\sigma$. The particles were placed in a cubic box, and  NEMD simulations were performed using the SLLOD equations of motion. Lees-Edwards shear boundary conditions were applied to eliminate effects of surfaces and of the small system volume. A Gaussian isokinetic thermostat was used to keep the temperature constant. The equations of motion were integrated using the operator-splitting algorithm of Pan {\it et al.}\cite{Pan/others:2005} implemented in the GPU-accelerated MD code RUMD\cite{rumd:2012}. While RUMD, like many GPU codes, uses mainly single-precision floating-point arithmetic, the summation of kinetic energy and similar quantities required for the isokinetic thermostat was done in double precision to avoid unacceptable numerical drift in the kinetic energy. 

For each reference state point of the SCLJ system we plotted the instantaneous virial versus the instantaneous potential energy. A linear regression to data according to Eq.\ \eqref{gamma1} gave $\gamma=5.75$ and the correlation coefficient $R=0.96$ at the zero-strain-rate reference state point. Within statistical errors the value of $\gamma$ was found to be independent of strain rate at the reference density and temperature ($\rho_0=0.84$, $T_0=0.8$), while the correlation coefficient increases with increasing strain rate, up to about 0.99 at $\dot{\gamma}=2.5$; at higher values of $\dot{\gamma}$ the system enters the so-called string phase, an artifact of the thermostat\cite{string}, and both $R$ and $\gamma$ drop significantly. The values of $R$ show that the system is simple in the Roskilde sense of the term, i.e., has good isomorphs. We studied in detail a single isomorph obtained by increasing the density by 5, 10, and 15 percent with respect to $\rho_0$; the new temperatures and strain rates corresponding to each new density were calculated using Eqs.\ \eqref{T} and \eqref{gamma}, respectively, after first determining $h(\tilde\rho)$ via Eq. \eqref{H} from simulations at the reference state point. Note that the strain rate independence of $\gamma$ has a non-trivial consequence: the projected isomorphs
coincide for different strain rates -- and coincide with the equilibrium isomorph. Therefore, having this isomorph at one reduced strain rate, one can generate isomorphs and different reduced strain rates using the same $(\rho, T)$ values.

The same procedure was applied to the KABLJ system ($800$ particles of type A and $200$ particles of type B interacting via Lennard-Jones pair potentials in a cubic box) with reference state points given by $\rho_0=1.2$, $T_0=0.579$ (in LJ units referring to the A particle parameters), and $\dot{\gamma}\leq1.2$. The KABLJ potential parameters are as follows: $\sigma_{AA}=1$ $\sigma_{AB}=0.8$, $\sigma_{BB}=0.88$, $\varepsilon_{AA}=1$, $\varepsilon_{AB}=1.5$, $\varepsilon_{BB}=0.5$, $m_{B}=m_{A}$. We here used the standard cut-off radius $2.5 \sigma_{AA}$. From the linear fit of instantaneous virial versus instantaneous potential energy the values $\gamma= 5.17$ and $R=0.94$ were determined for zero strain rate at the reference density. By using this value of $\gamma$ and Eqs. \eqref{H} and \eqref{T} one can go from one state point to another isomorphic state point. We studied a single isomorph, changing density by $\pm 5 ,+10$, and $+15$ percent with respect to $\rho_0$. In the KABLJ system, like in the SCLJ system, $\gamma$ was independent of shear rate and $R$ increased with increasing shear rate. 

For both systems we let the system go to its stationary state by running without output for some time (of order $10^5$ time steps). The production phase involved of order $10^7$-$10^8$ time steps, depending on the strain rate. This allowed for an accurate determination of structural and time-dependent correlation functions, as well as of the viscosity.

\section{Simulation results}\label{sim_results}

\begin{figure}
\includegraphics[width = 0.45 \textwidth]{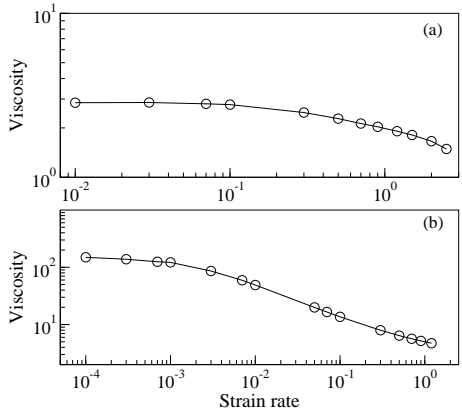}
\caption{Viscosity as a function of strain rate for (a) the SCLJ (single-component Lennard-Jones) system at $\rho=0.84$, $T=0.8$, and (b) the KABLJ (Kob-Andersen binary Lennard-Jones) system at $\rho =1.2$, $T=0.579$. The transition to the nonlinear regimes occurs around $\dot{\gamma} \sim 0.6$ for SCLJ and around $\dot{\gamma} \sim 0.002$ for KABLJ.} \label{vis}
\end{figure}

In fluids undergoing planar Couette flow, the viscosity $\eta$ gives the response to the applied flow. Figure \ref{vis} shows viscosity versus strain rate for both systems studied. At low strain rates the viscosity is constant (linear behavior), but at higher strain rates it starts to decrease. This ``shear thinning'' is well known from experimental rheology of, e.g., polymeric liquids\cite{fdr,bair,thinning,kim}. We find the onset of the transition at $\dot{\gamma}\sim 0.6$ for SCLJ and  $\dot{\gamma}\sim 0.002$ for KABLJ at the reference state point. 

\begin{figure}
\includegraphics[width = 0.4 \textwidth]{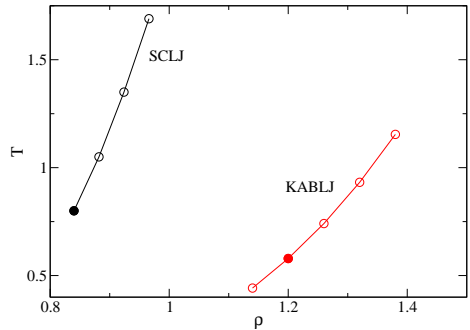}
\caption{Density-temperature phase diagram showing four isomorphic state points of the SCLJ system and five for the KABLJ  system (i.e., the projected isomorphs). The reference state points are marked with full symbols.} \label{phase}
\end{figure}

The procedure explained in Sec. \ref{generating_isom} was used to generate isomorphic state points for both systems, starting from the reference state points. Figure \ref{phase} shows the isomorphic state points' densities and temperatures.

\begin{figure}
\includegraphics[width = 0.4 \textwidth]{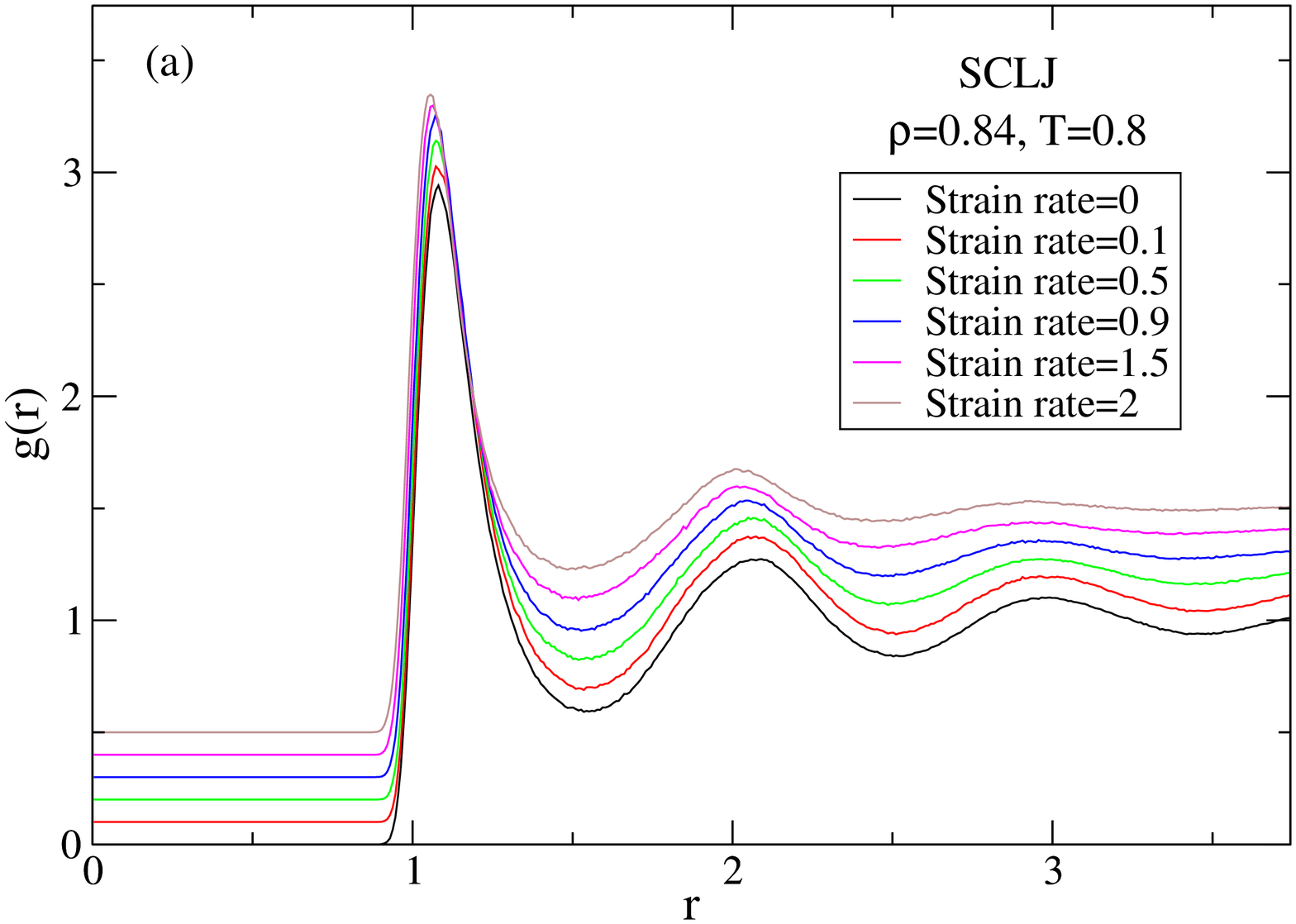}
\includegraphics[width = 0.4 \textwidth]{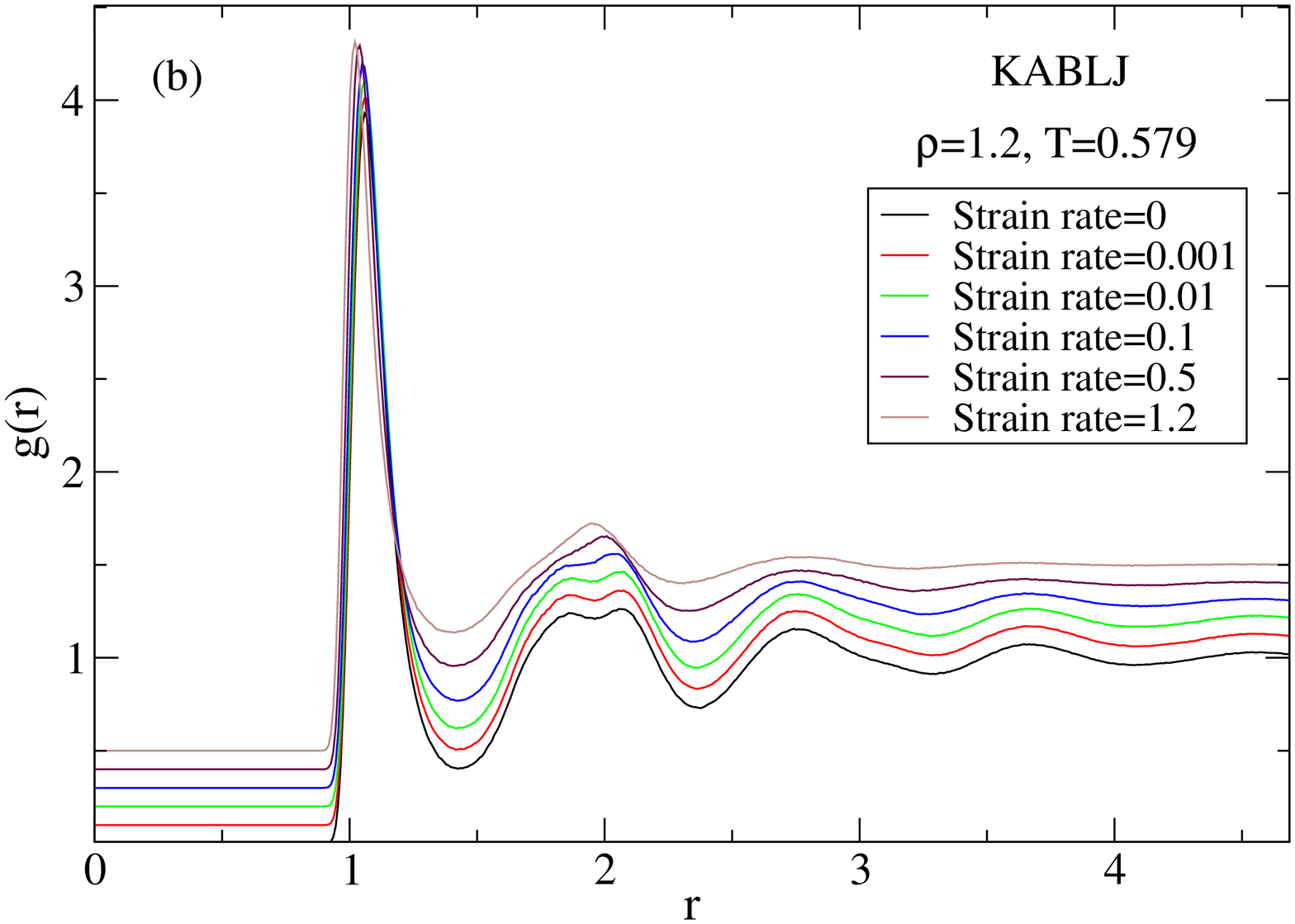}
\caption{Radial distribution function $g(r)$ of (a) the SCLJ system and (b) the KABLJ system at the reference state points with different strain rates. For clarity the radial distribution functions have been displaced by $0.1n$ with $n=0,...,5$. For the SCLJ system there is a change of structure between strain rate 0.5 and 0.9, consistent with the onset of shear thinning. The same structure change takes place for the KABLJ system somewhat above the onset of shear thinning.} \label{fig1}
\end{figure}

Figure \ref{fig1}(a) shows the radial distribution function of the SCLJ system at $\rho_0=0.84,T_0=0.8$, for selected strain rates below $2$. Figure \ref{fig1}(b) shows the same function for the KABLJ system at $\rho_0=1.2,T_0=0.579$,  with strain rates below $1.2$. For both systems the structure changes once the strain rate exceeds the value corresponding to the transition to nonlinear behavior. As mentioned earlier, the so-called string phases appear at shear rates higher than those presented here, an artifact of the thermostat\cite{string}.

\begin{figure}
\includegraphics[width = 0.4 \textwidth]{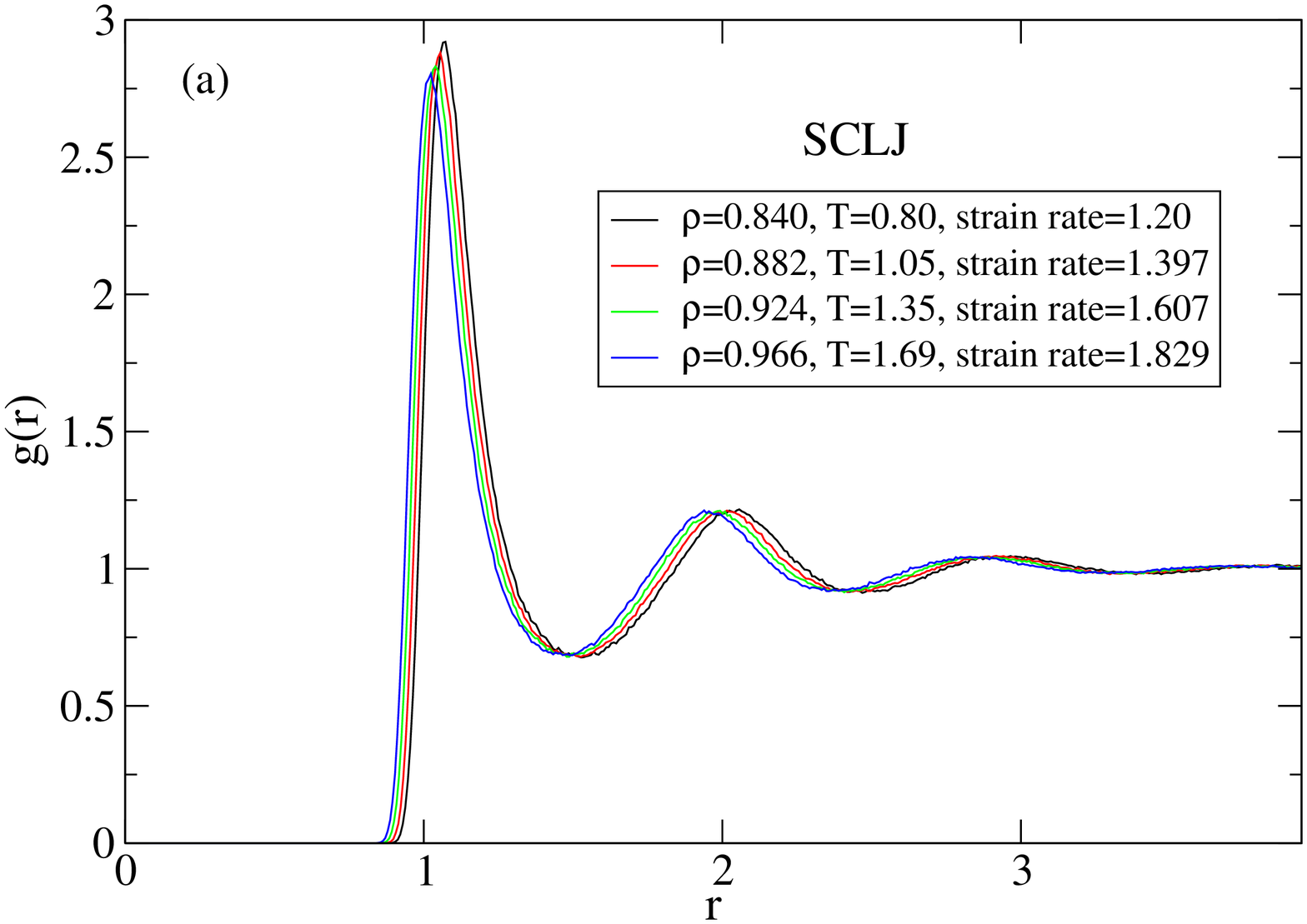}
\includegraphics[width = 0.4 \textwidth]{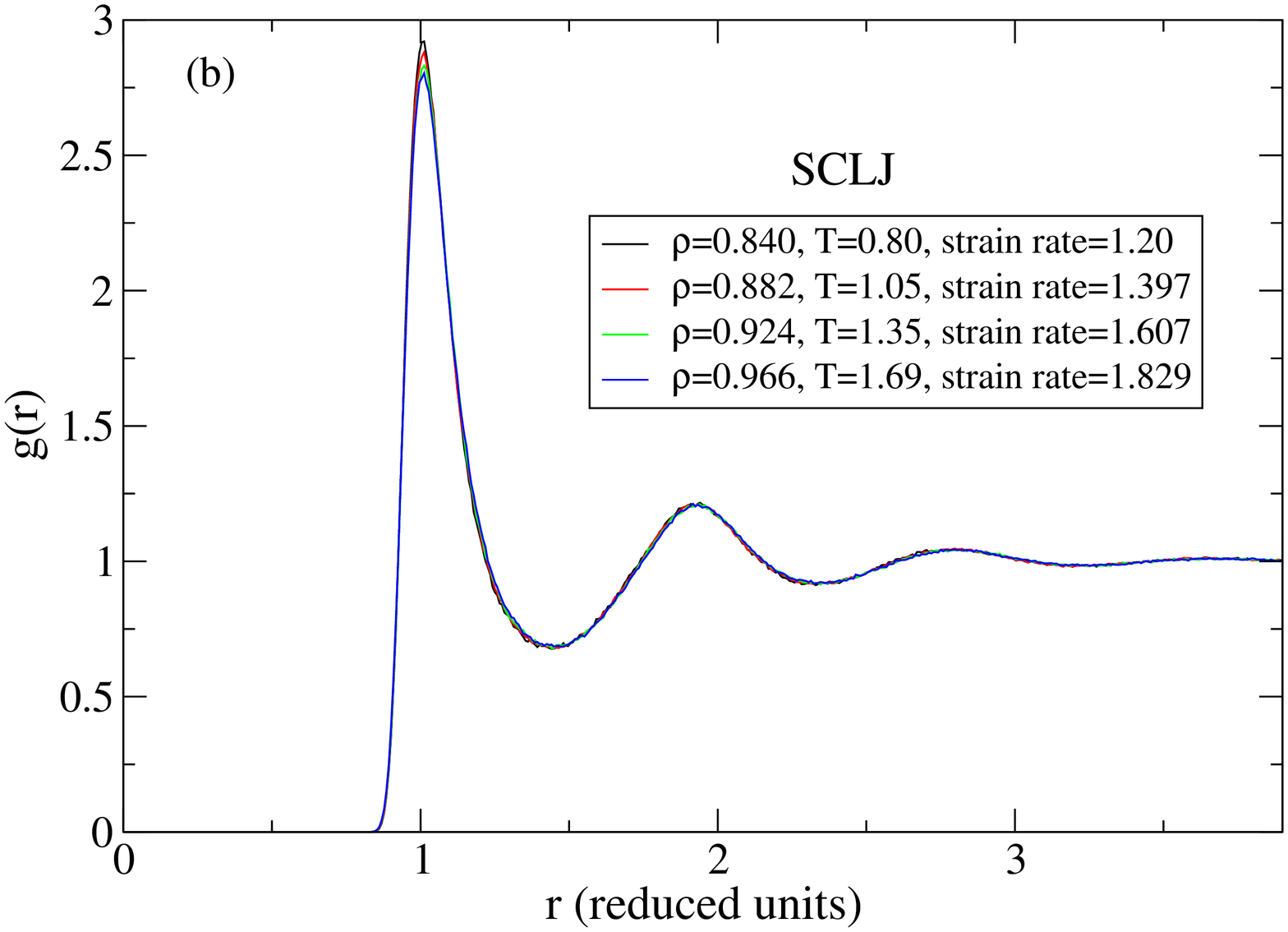}
\includegraphics[width = 0.4 \textwidth]{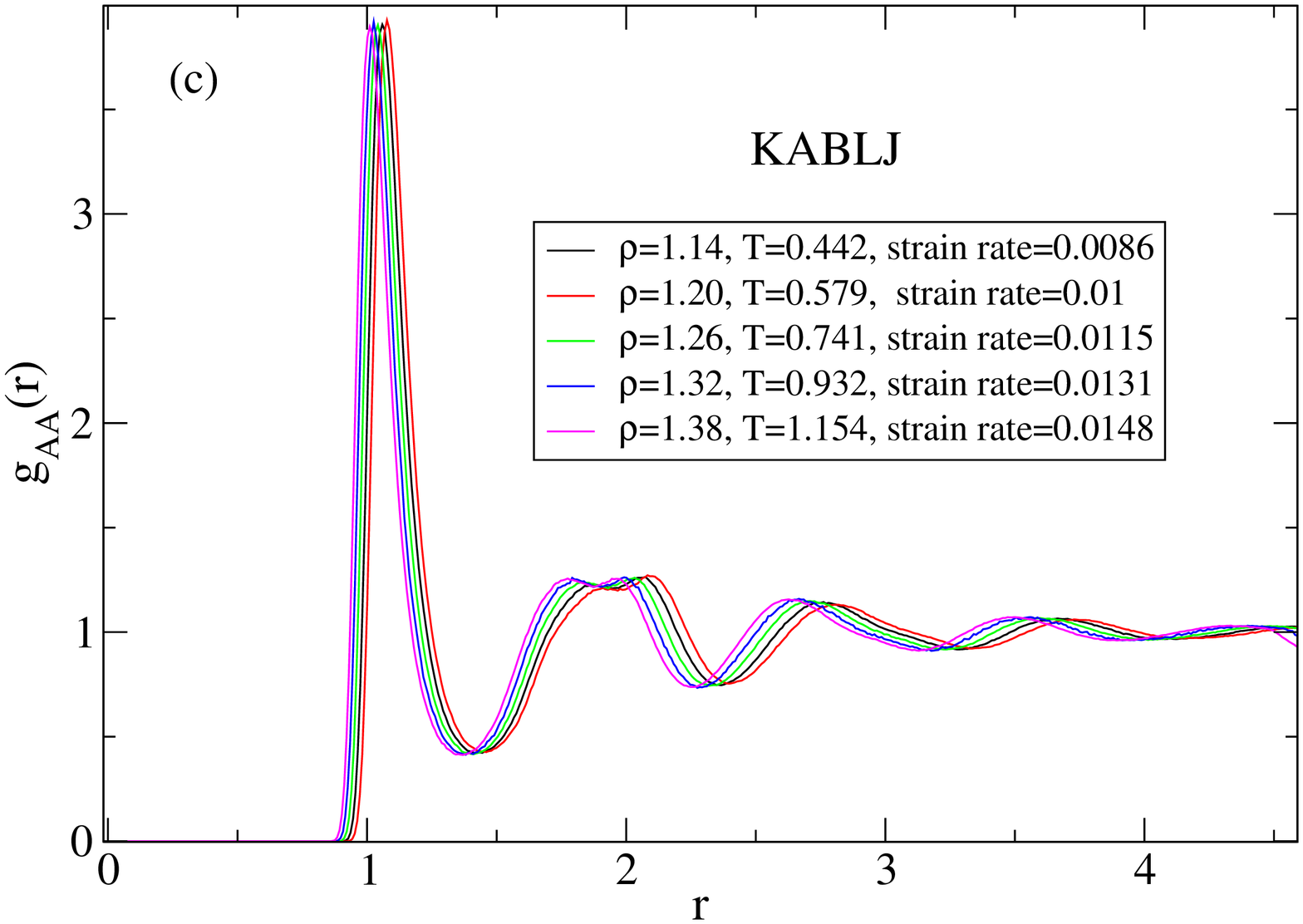}
\includegraphics[width = 0.4 \textwidth]{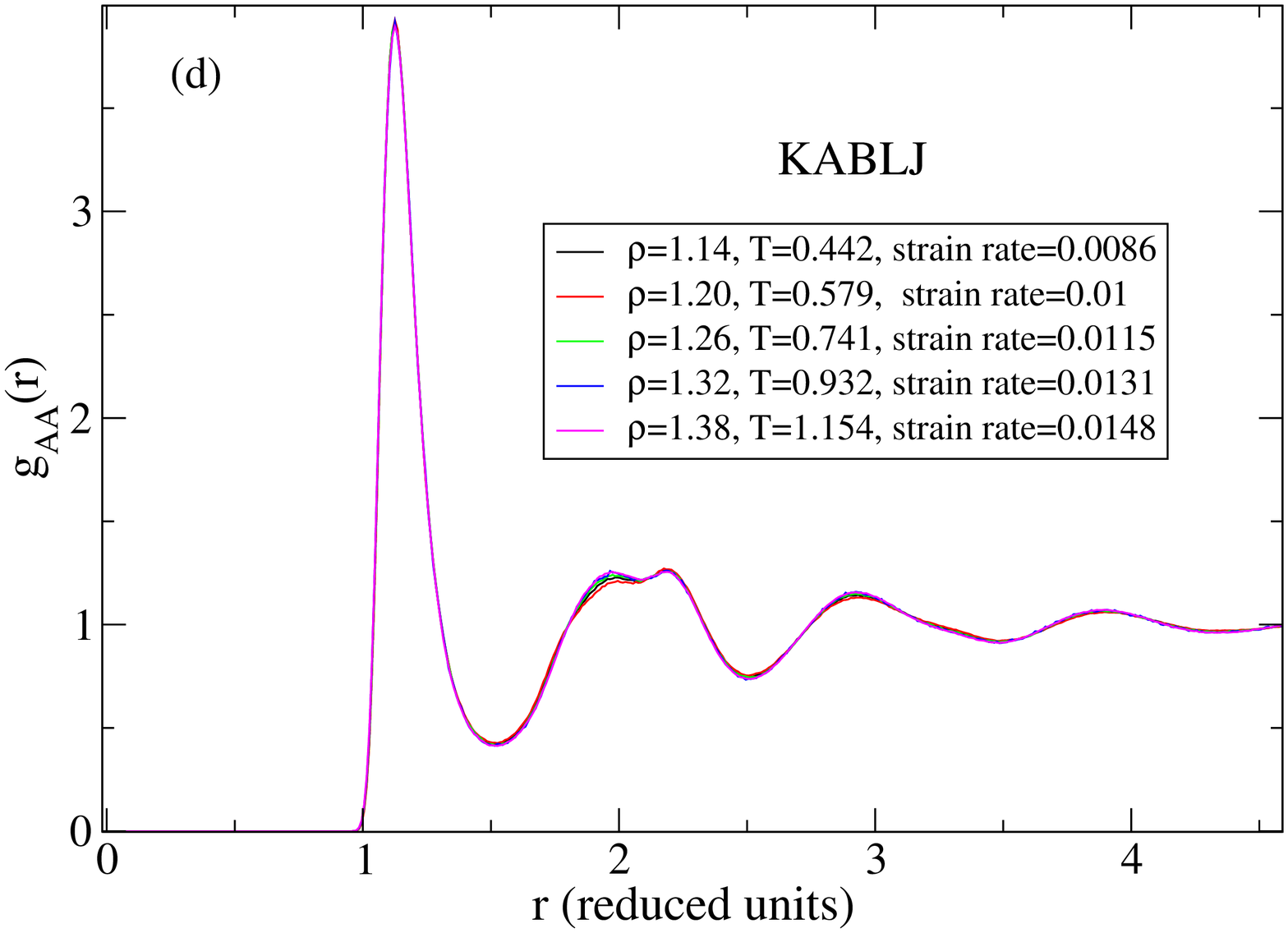}
\caption{Radial distribution function for the four isomorphic state points of the SCLJ system in (a) non-reduced units and (b) reduced units. (c) and (d) Radial distribution function of the A particles for the five isomorphic state points of the KABLJ system in (c) non-reduced and (d) reduced units. To a good approximation the structure is invariant along the isomorphs.} \label{fig2}
\end{figure}

Figure \ref{fig2}(a) shows the radial distribution functions of four isomorphic state points of the SCLJ system at a reduced strain rate corresponding to nonlinear flow. In Fig.~\ref{fig2}(b) $g(r)$ is plotted as a function of reduced distance, $\tilde{r}\equiv\rho^{1/3}r$. The good collapse of curves confirms the invariance of structure. The same result was obtained for the KABLJ system; Figs.~\ref{fig2}(c) and (d) show the radial distribution function for the five generated isomorphic state points in non-reduced and reduced units for a reduced strain rate in the nonlinear regime. 

\begin{figure}
\includegraphics[width = 0.4 \textwidth]{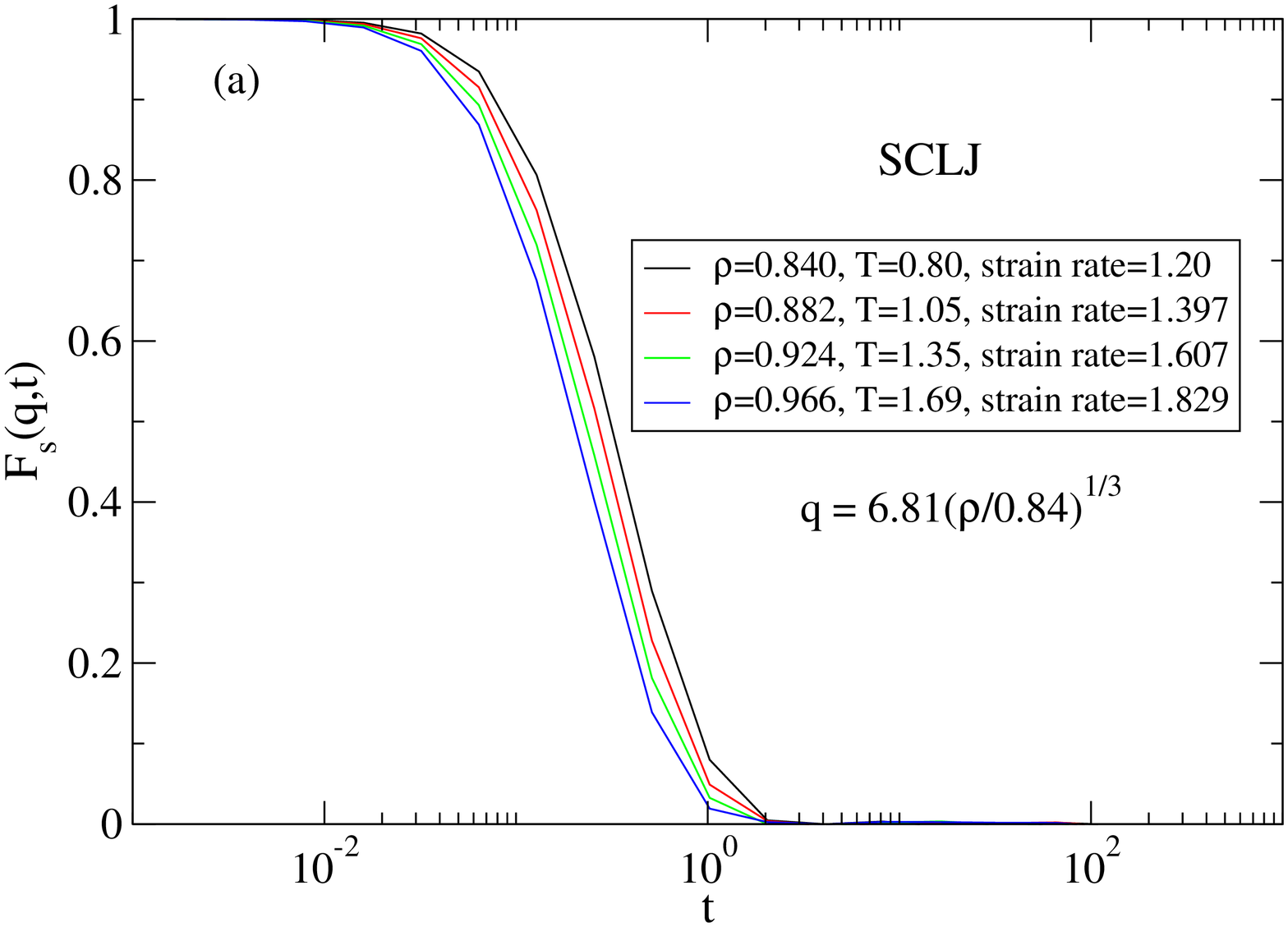}
\includegraphics[width = 0.4 \textwidth]{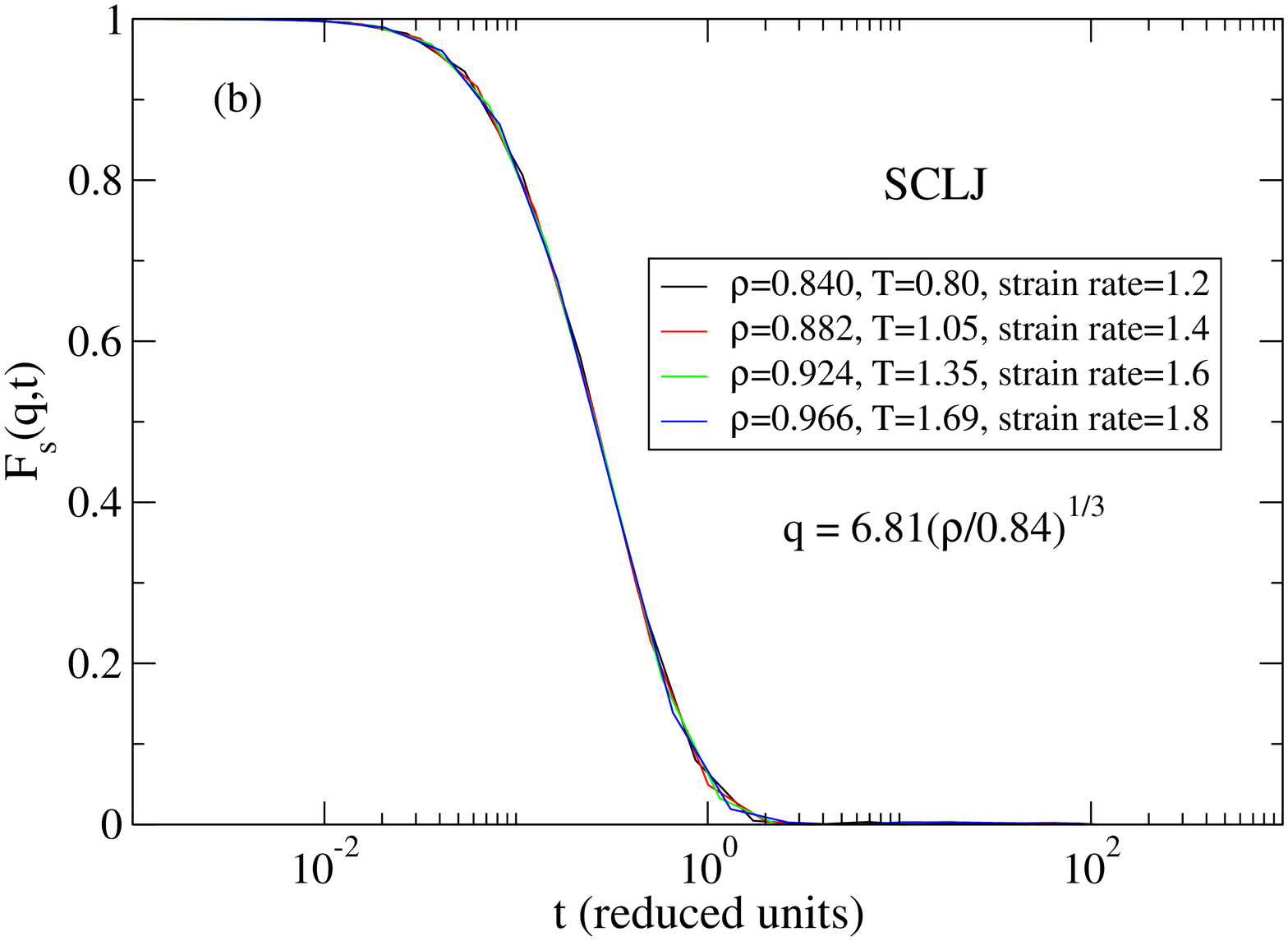}
\includegraphics[width = 0.4 \textwidth]{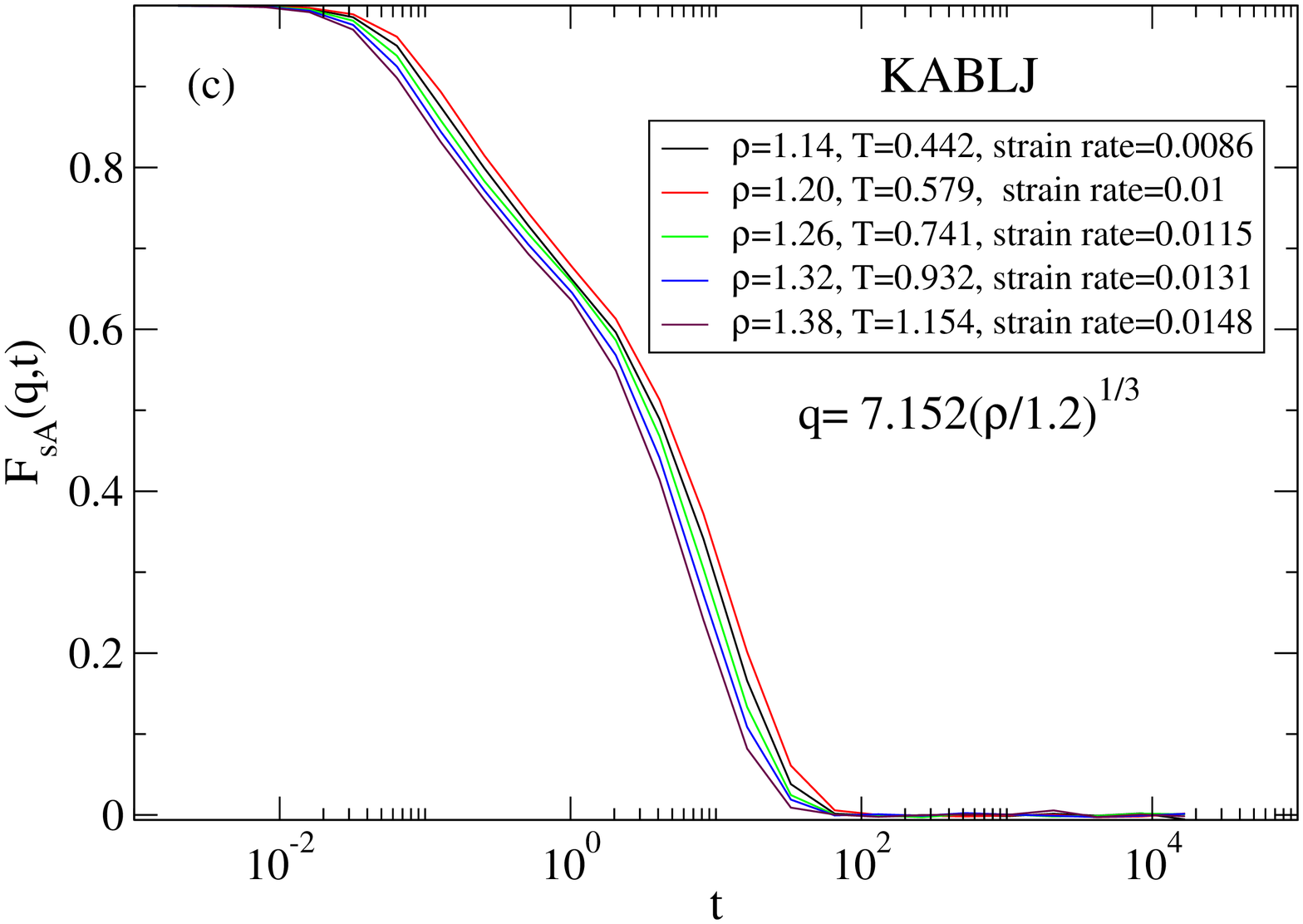}
\includegraphics[width = 0.4 \textwidth]{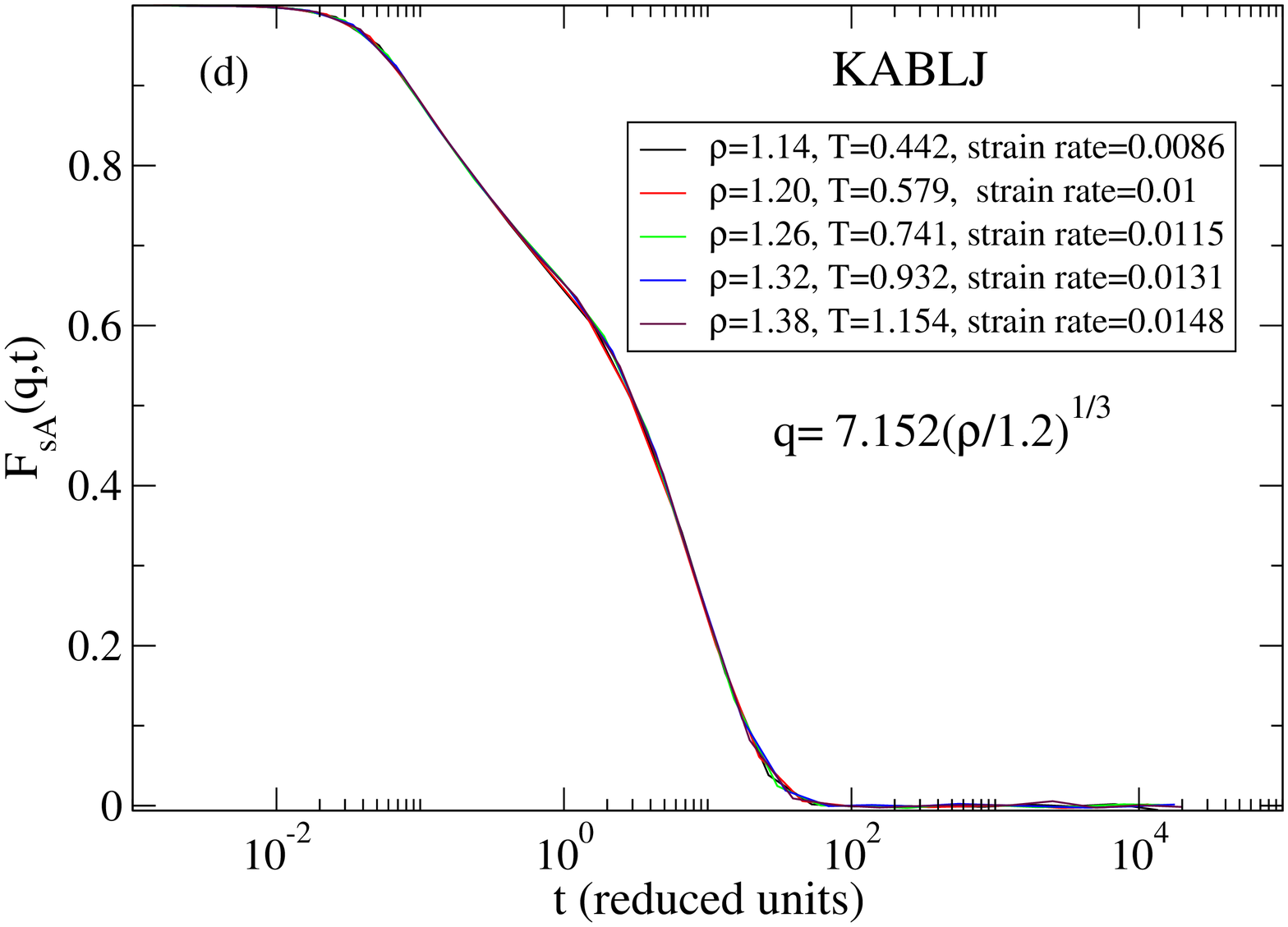}
\caption{Intermediate scattering function (transverse displacements) for the four isomorphic state points of the SCLJ system at $q=6.81(\rho/0.84)^{1/3}$ as a function of (a) ordinary time and (b) reduced time. The next two figures show intermediate scattering function (A particles, transverse displacements) for the five isomorphic steady state points of the KABLJ system at $q=7.152(\rho/1.2)^{1/3}$ as a function of (c) ordinary time and (d) reduced time. The collapses in (b) and (d) demonstrate isomorph invariance of the dynamics in reduced units.}\label{fig3}
\end{figure}

To investigate the dynamical invariance of isomorphic state points we calculated the intermediate scattering function $F_s(q,t)$.  In the presence of a flow, rather than attempting to disentangle the stochastic deviations from the average flow it is convenient to consider only displacements transverse to the flow direction. Following tradition we chose a $q$-value close the first peak of the static structure factor. Along an isomorph this $q$ scales as $\rho^{1/3}$, and a correct comparison in reduced units must take this into account. We chose $q=6.81$ at the reference density for the SCLJ system and $q=7.152$ at the reference density for the KABLJ system. Figure~\ref{fig3}(a) shows the transverse $F_s(q,t)$ for the SCLJ system as a function of ordinary time (but scaled $q$), while Fig.~\ref{fig3}(b) shows the same quantity as a function of reduced time.  Figures~\ref{fig3}(c) and (d) show the corresponding results for the KABLJ system. A good collapse of curves is seen when reduced time units are used, showing that the dynamics are invariant along the isomorphs.

\begin{figure}
\includegraphics[width = 0.4 \textwidth]{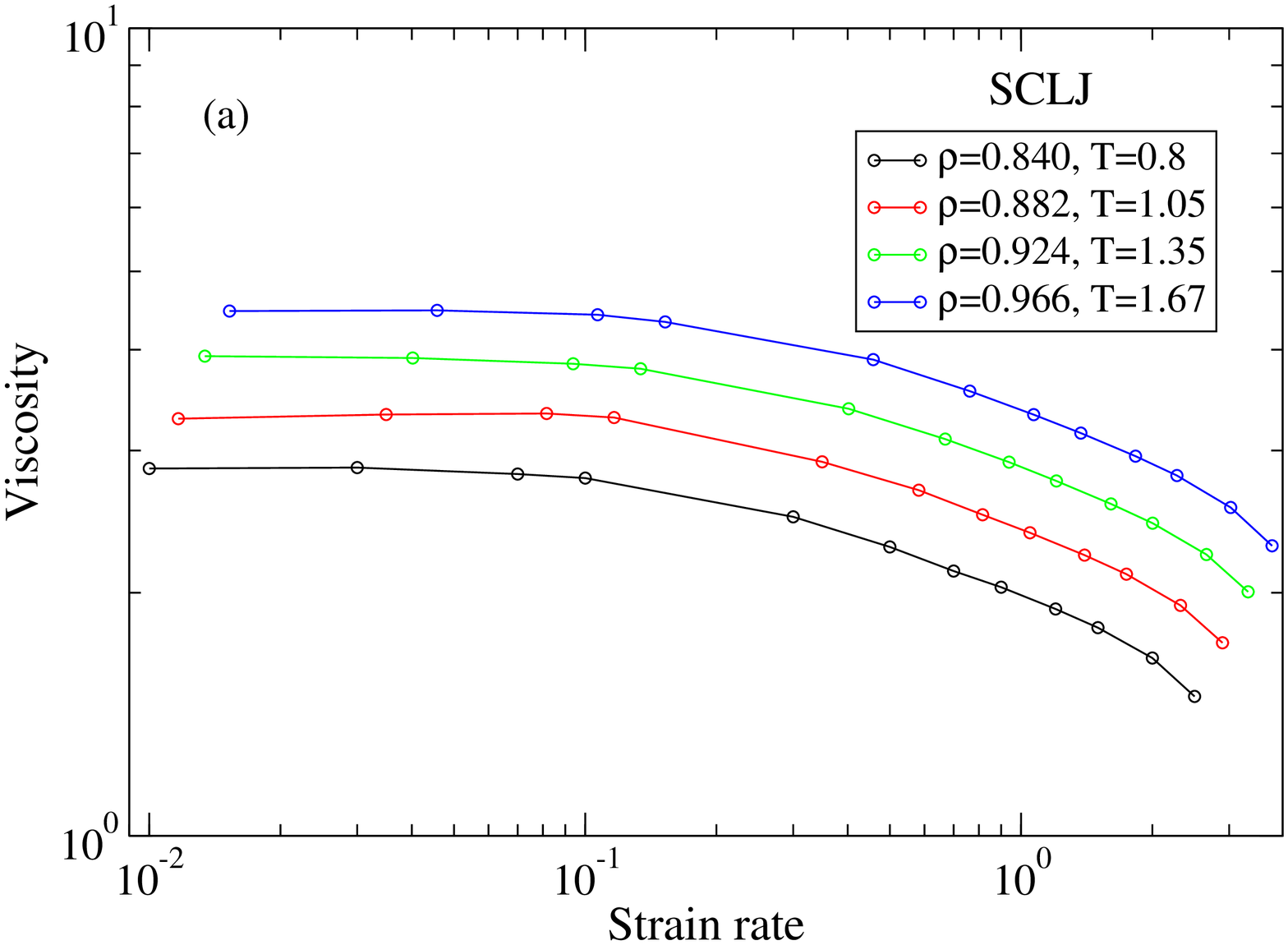}
\includegraphics[width = 0.4 \textwidth]{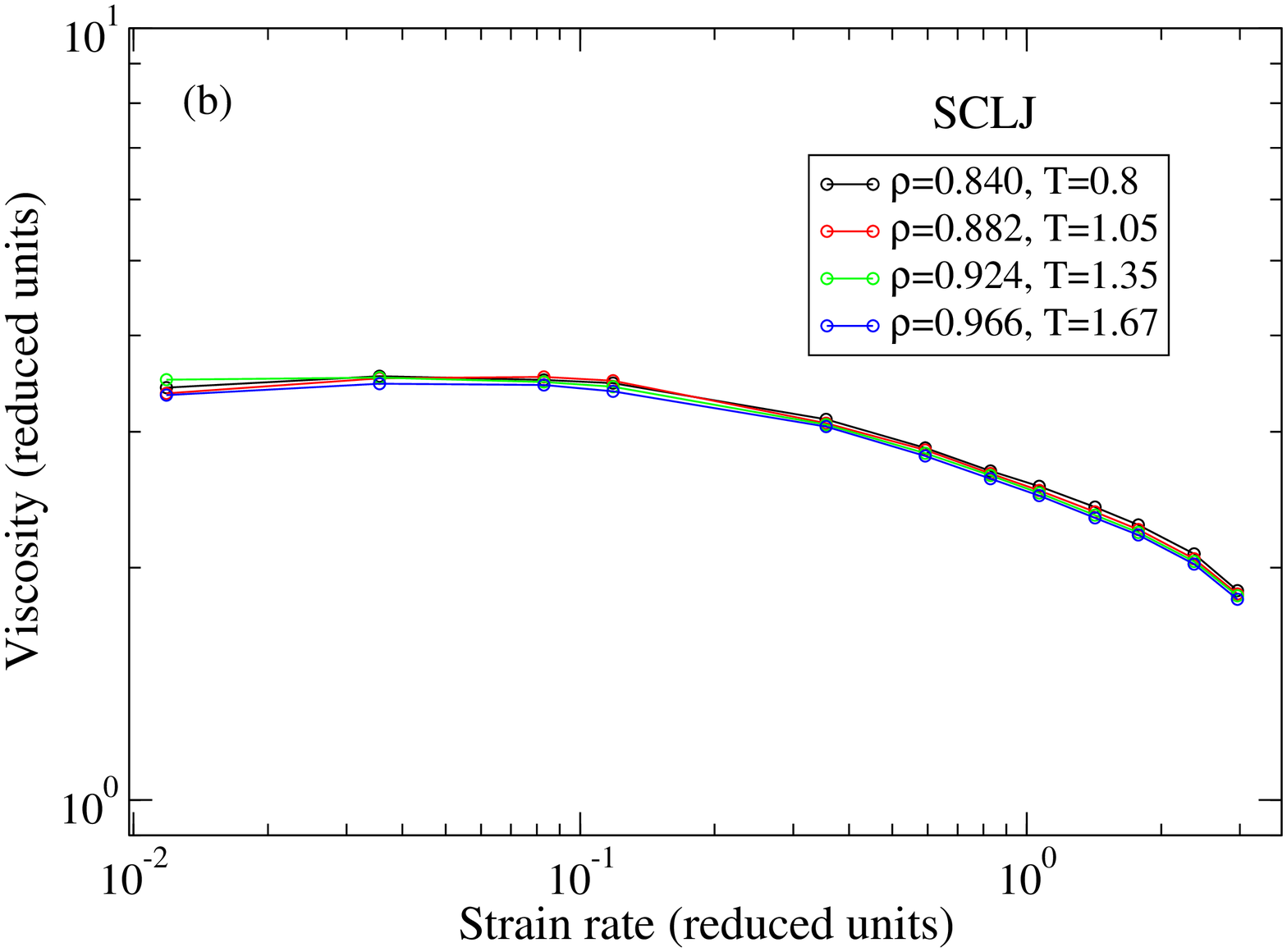}
\includegraphics[width = 0.4 \textwidth]{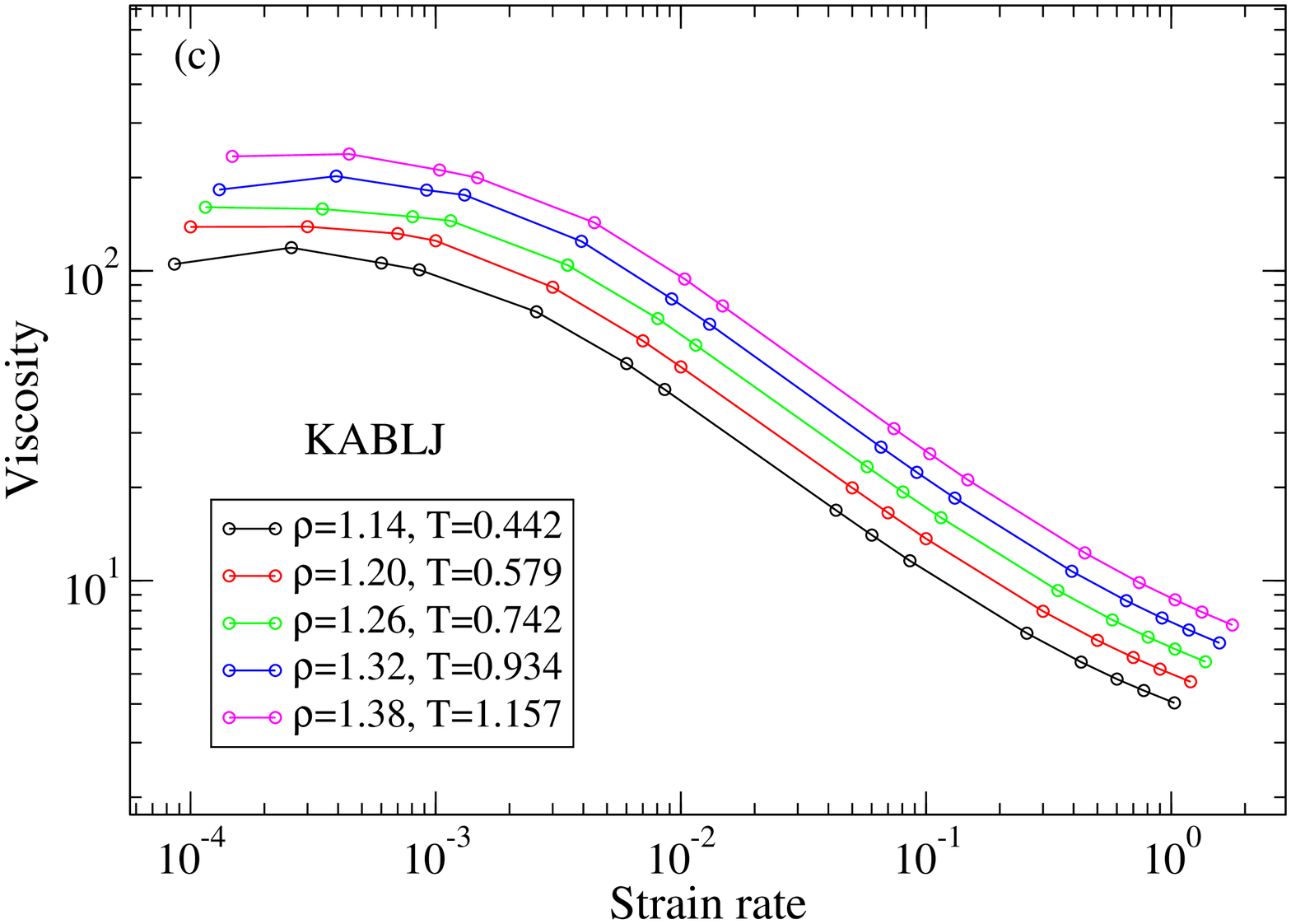}
\includegraphics[width = 0.4\textwidth]{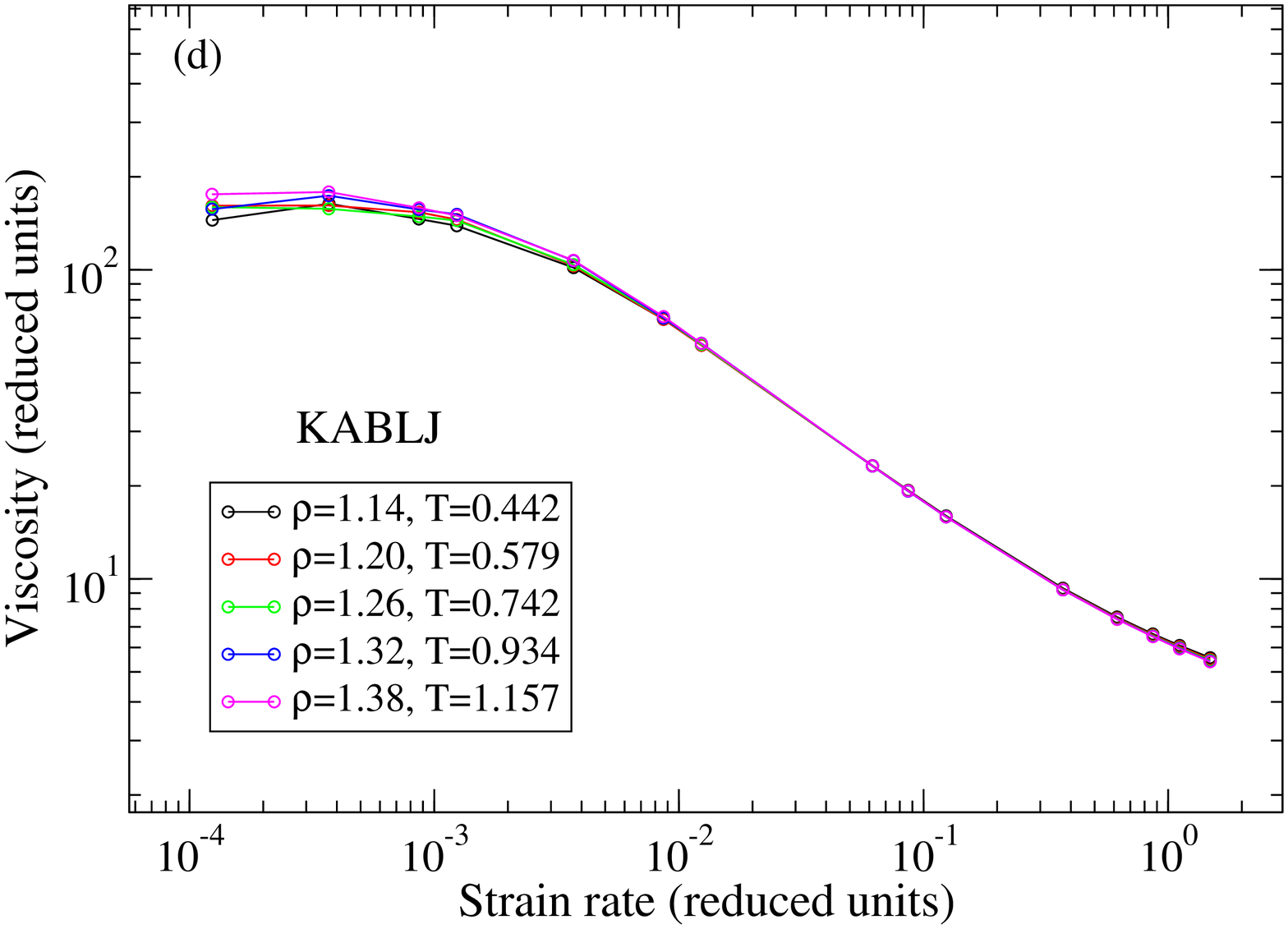}
\caption{(a) Viscosity versus strain rate for the SCLJ system at the four points shown in Fig.~\ref{phase}; (b) reduced viscosity $\widetilde{\eta}=\eta/(\rho^{2/3}T^{1/2})$ versus reduced strain rate for the same state points. (c) Viscosity versus strain rate for the five state points of the KABLJ system shown in Fig.~\ref{phase}; (d) $\widetilde{\eta}$ versus reduced strain rate for the same state points.}\label{fig6}
\end{figure}

We now consider rheology. According to the isomorph theory transport quantities such as the reduced diffusion coefficient and the reduced viscosity are invariant along an isomorph. Rheology can be said largely to be concerned with strain-rate dependence, so now we include data from a range of strain rates. As explained in Section~\ref{model_sim}, since the projected isomorph is independent of the strain rate, the $(\rho,T)$ values from the starting isomorph can be used. Simulations were run for a range of strain rates, thus generating data for a whole family of isomorphs parameterized by reduced strain rate. Figures~\ref{fig6}(a) and (c) show the viscosity of the SCLJ and KABLJ systems as functions of strain rate for different $(\rho,T)$ points along the common projected isomorph. The viscosity decreases upon increasing the strain rate, which is the already mentioned shear thinning effect\cite{fdr,bair,thinning,kim}. In Figs.~\ref{fig6}(b) and (d) the reduced viscosity $\widetilde{\eta}\equiv\eta/(\rho^{2/3}T^{1/2})$ is plotted as a function of reduced strain rate. The collapse of the curves demonstrates isomorph invariance of the shear-thinning behavior and confirms that the projected isomorphs coincide.

We also simulated thermodynamic quantities, focusing on potential energy and pressure. These quantities are not inherently isomorph invariant\cite{IV}. However, based on the argument\cite{II} that strong correlations and the existence of isomorphs in non-IPL (inverse power law) potential systems is linked to a decomposition of the pair potential into an effective IPL part plus an almost linear part, one expects the quantity $g(Q)$ of Eq. \eqref{isom} to depend mainly on density and negligibly on temperature and strain rate: The sum of all pair energies from the linear part of the pair potential is roughly constant at a given state point, and depends mainly on volume when different state points are considered\cite{II}. The dependence on volume explains why quantities such as energy, free energy, and pressure are not isomorph invariant. Under the assumption that $g(Q)$ does not depend on strain rate, the isomorph theory predicts that the strain-rate dependent parts of potential energy and virial, $U(\rho, T,\dot{ \gamma} )- U(\rho, T, 0)$ and $W(\rho, T, \dot{\gamma}) -W(\rho, T, 0)$, are both isomorph invariant when given in reduced units. Notice that the same must be true for the total energy and pressure since the subtraction eliminates the kinetic terms.

\begin{figure}
\includegraphics[width = 0.4 \textwidth]{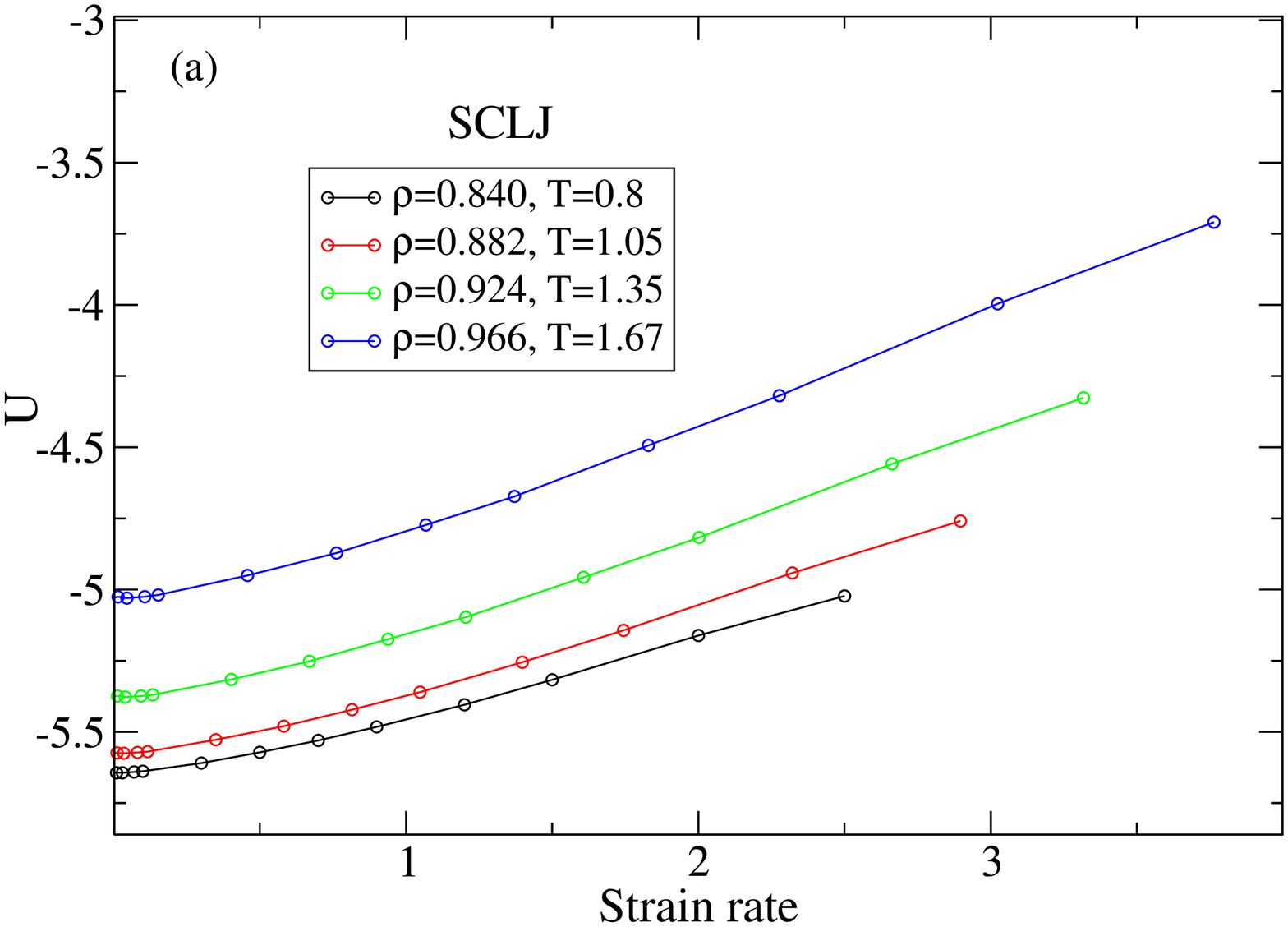}
\includegraphics[width = 0.4 \textwidth]{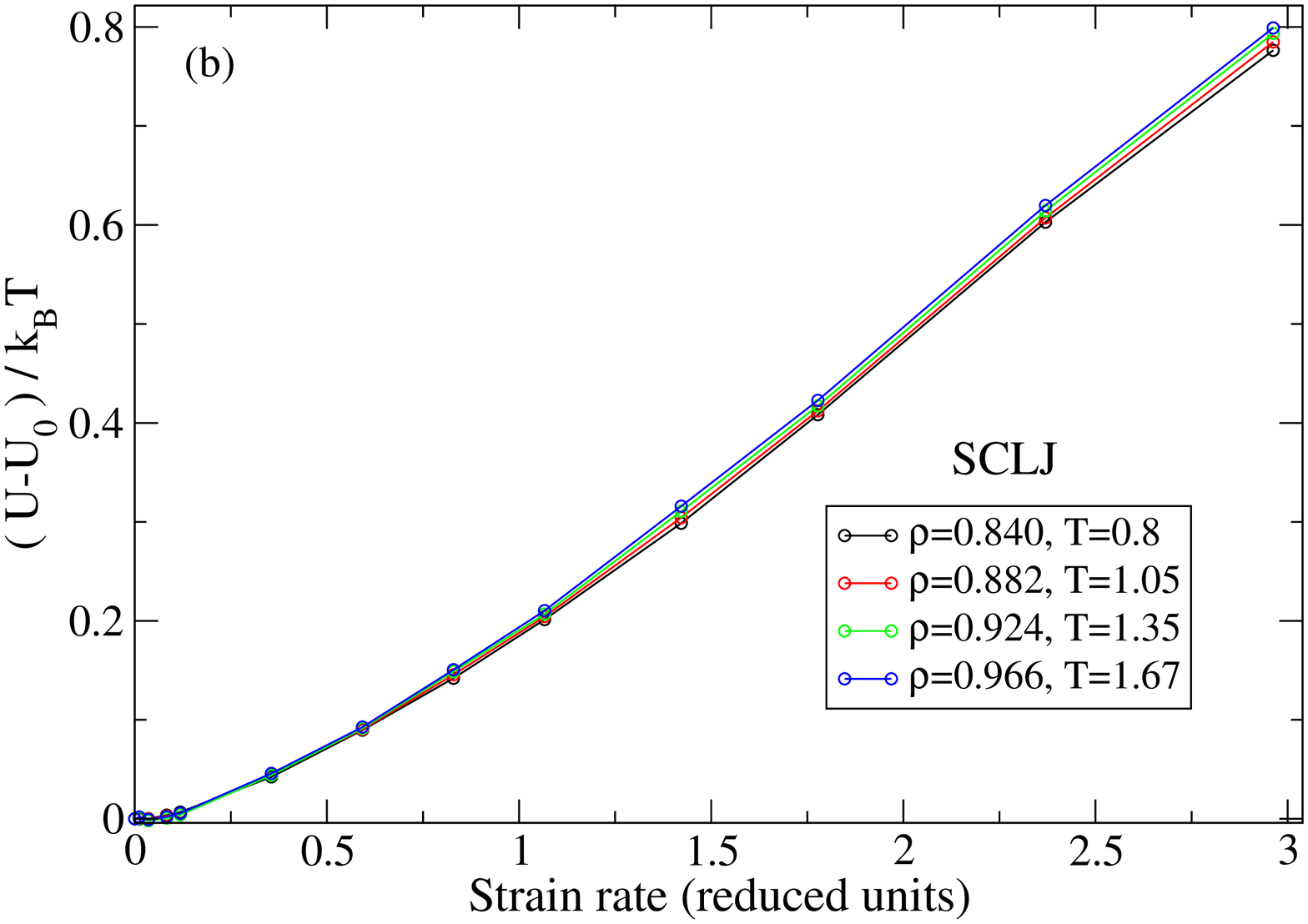}
\includegraphics[width = 0.4 \textwidth]{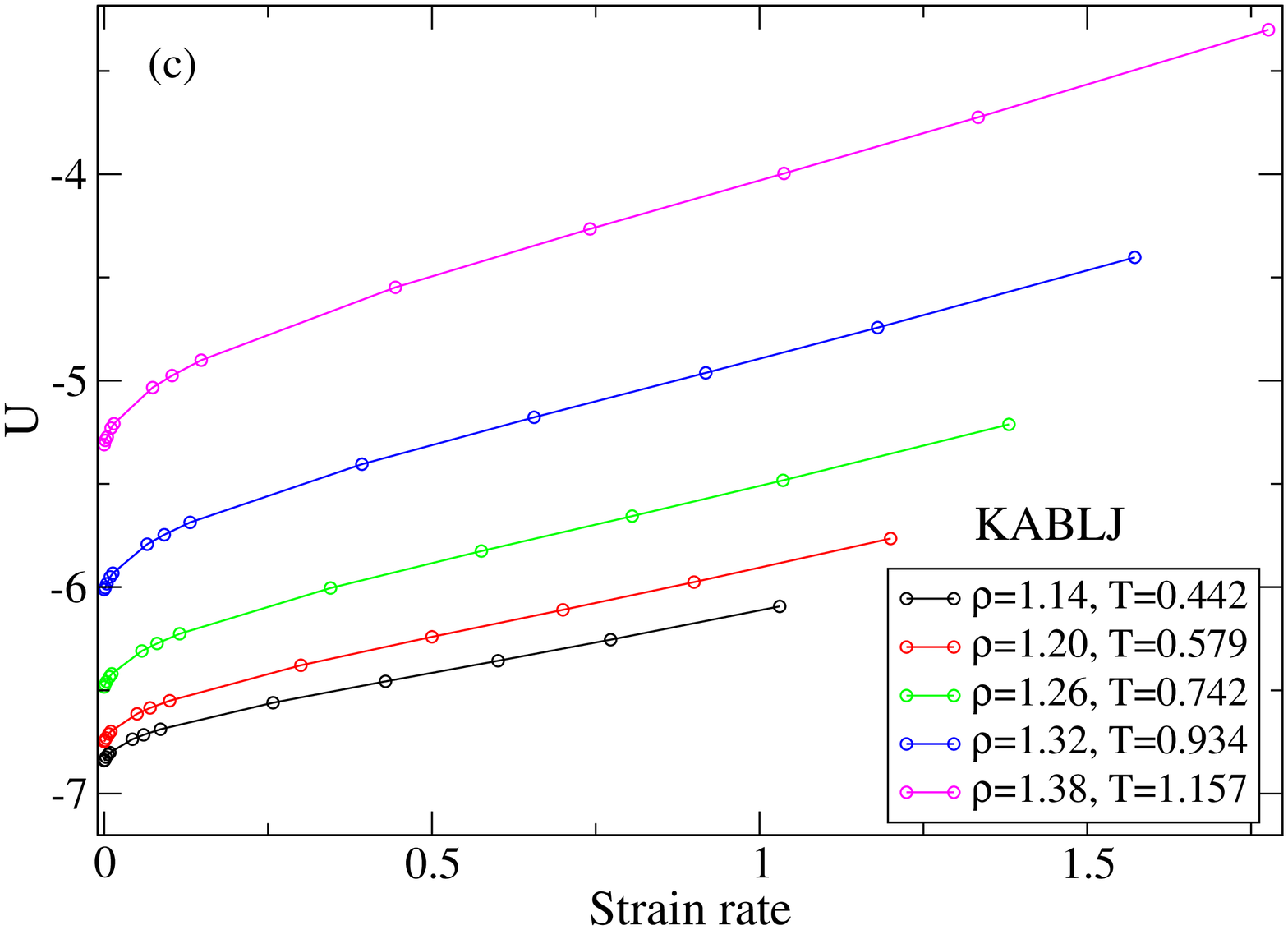}
\includegraphics[width = 0.4 \textwidth]{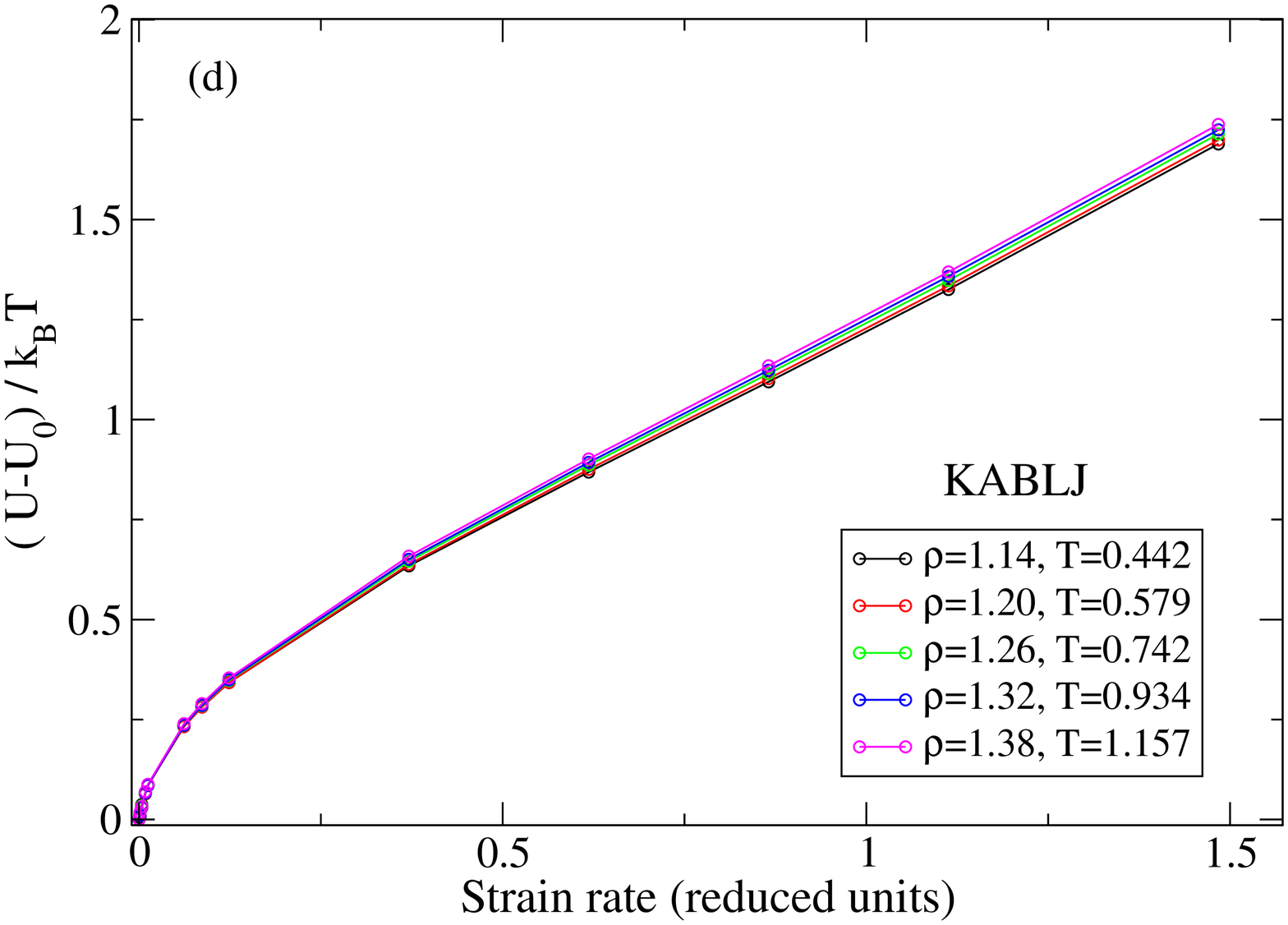}
\caption{(a) Potential energy versus strain rate for the SCLJ system at the four $(\rho,T)$ points shown in Fig.~\ref{phase}; (b) The strain-rate dependent reduced potential energy $(U-U_0)/k_{B}T$ versus reduced strain rate $t_{0}\dot{\gamma}$ where $U_0$ is the potential energy at zero strain rate. (c) Potential energy versus strain rate for the KABLJ system for the five $\rho,T$ points shown in Fig.~\ref{phase}; (d) $(U-U_0)/k_{B}T$ versus reduced strain rate.}\label{fig4}
\end{figure}

Figure~\ref{fig4}(a) shows the potential energy as a function of strain rate for the four SCLJ state points of the projected isomorph (Fig.~\ref{phase}). Figure \ref{fig4}(b) plots as a function of reduced strain rate the strain-rate dependent part of the reduced potential energy, $(U-U_0)/k_{B}T$, where $U$ is the (average) potential energy and $U_0$ the (average) potential energy of the zero-strain-rate isomorphic state point. The potential energy in non-reduced and reduced units for the KABLJ system is plotted in Figs.~\ref{fig4}(c) and (d), respectively. For both systems a good data collapse is seen.

\begin{figure}
\includegraphics[width = 0.4 \textwidth]{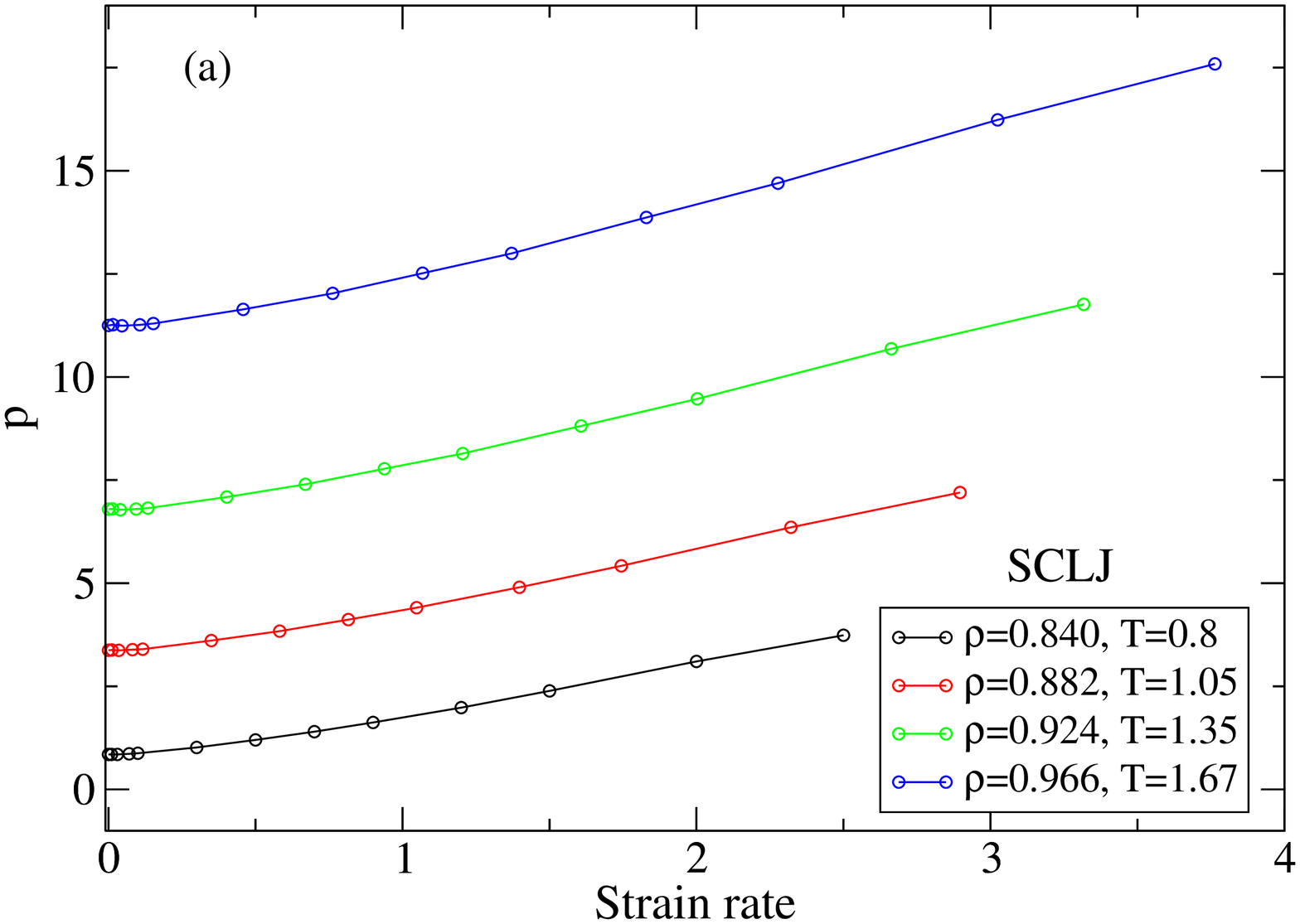}
\includegraphics[width = 0.4 \textwidth]{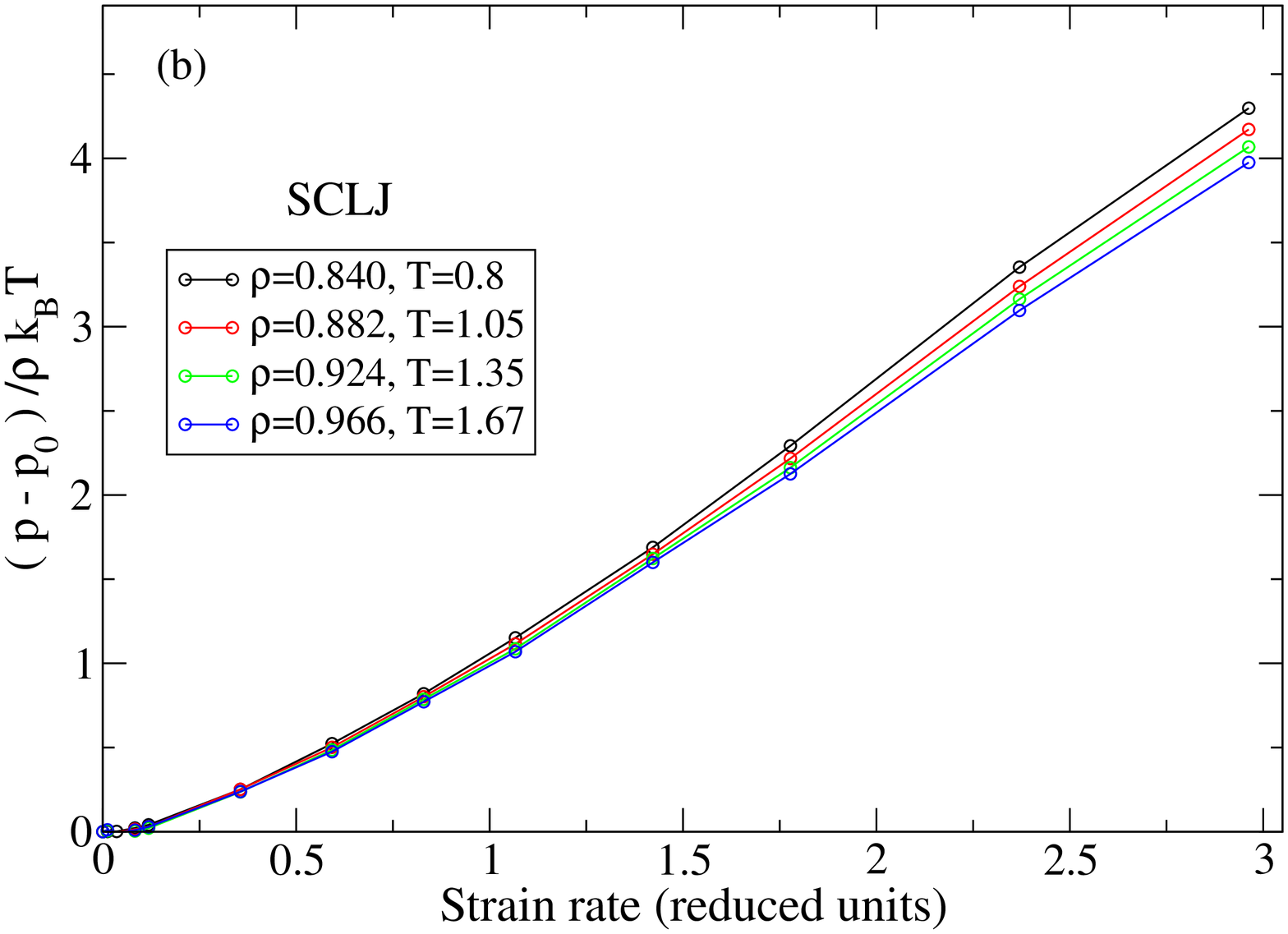}
\includegraphics[width = 0.4 \textwidth]{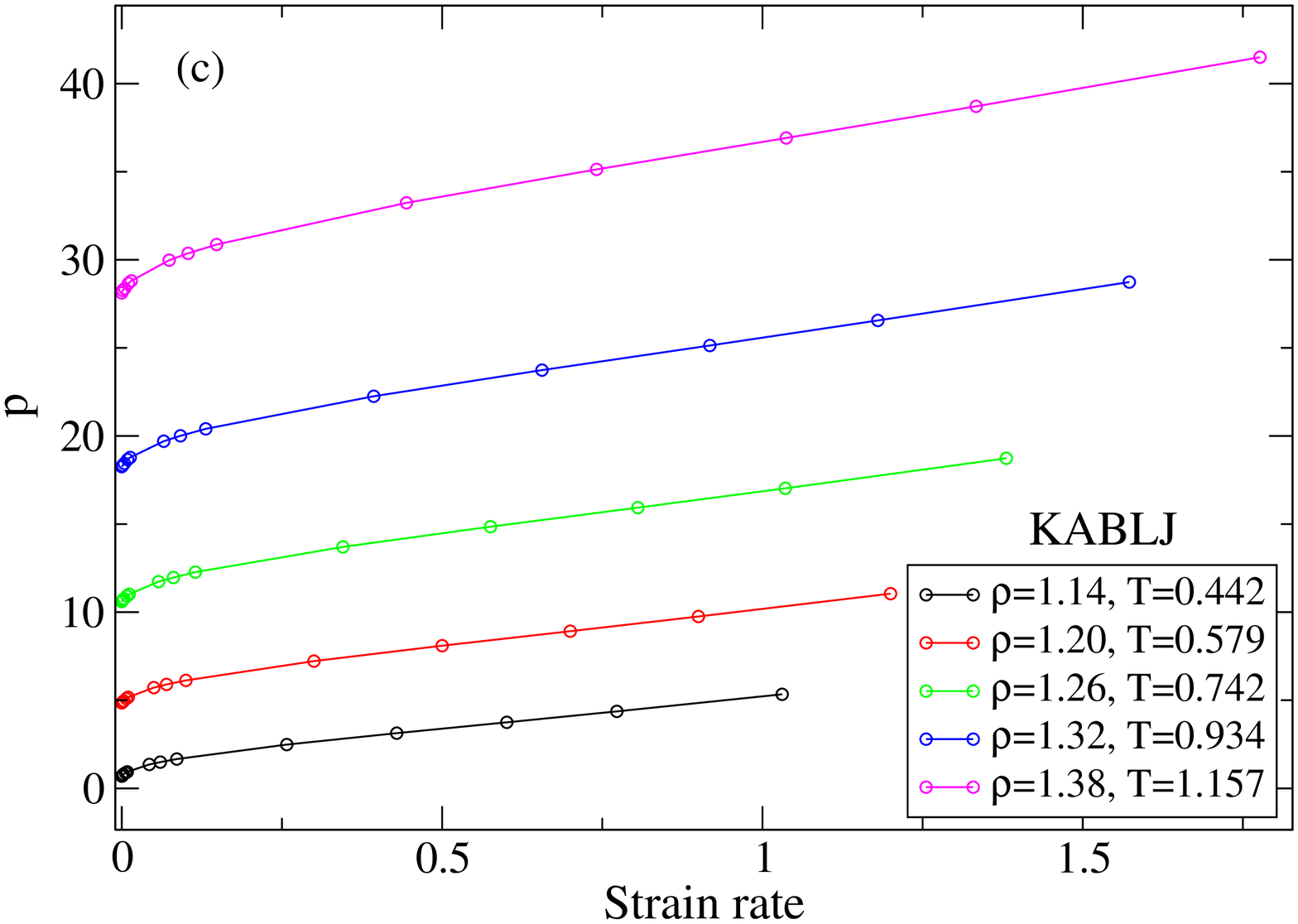}
\includegraphics[width = 0.4 \textwidth]{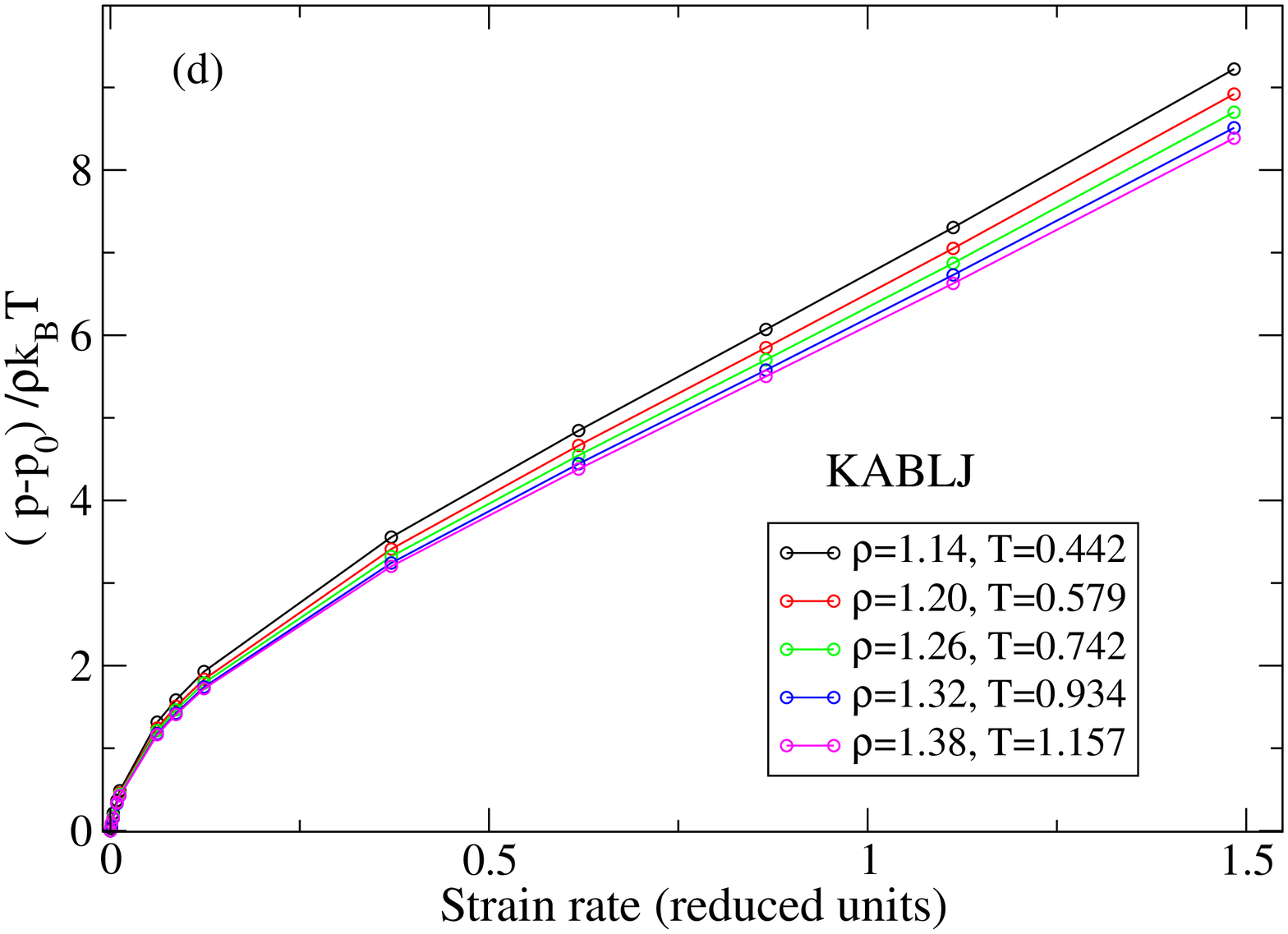}
\caption{(a) Pressure versus strain rate for the SCLJ system at the four state points of Fig.~\ref{phase}; (b) the strain-rate dependent reduced pressure $(p-p_0)/(\rho k_{B}T)$ versus reduced strain rate for the same state points. (c) Pressure versus strain rate for the KABLJ system at the five state points shown in Fig.~\ref{phase}; (d) $(p-p_0)/(\rho k_{B}T)$ versus reduced strain rate for the same state points.}\label{fig5}
\end{figure}

We did the same for pressure. Figures ~\ref{fig5}(a) and (c) show the pressure as a function of strain rate for the SCLJ and KABLJ systems, respectively. Figures~\ref{fig5}(b) and (d) show the corresponding strain-rate dependent reduced pressure $(p-p_0)/(\rho k_{B}T)$ as a function of reduced strain rate. The collapse is reasonable, but not as good as for the potential energy. We do not have an explanation of this.

\begin{figure}
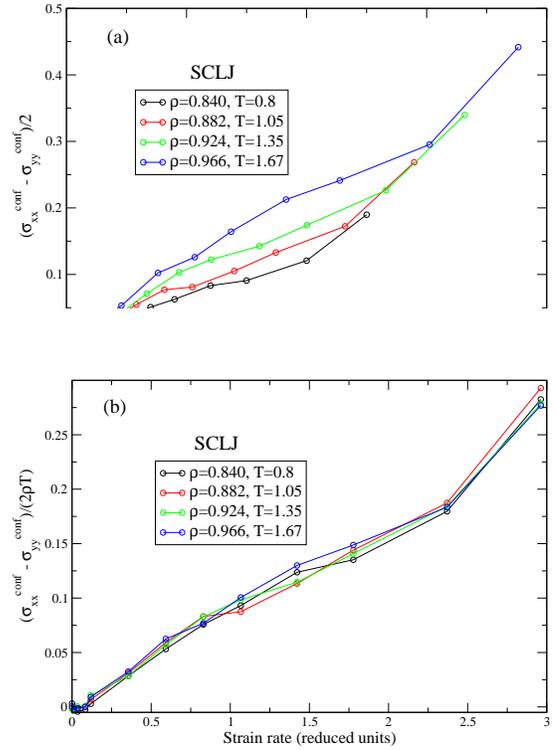

\includegraphics[width = 0.4 \textwidth]{NormalStsDiffs}
\includegraphics[width = 0.4 \textwidth]{NormalStsDiffsRed}
\caption{ Configurational parts of normal stress difference $(\sigma_{xx}-\sigma_{yy})/2$ for SCLJ in (a) normal units and (b) reduced units. While there is some statistical noise due to the inherent problems with subtracting similar quantities, there is a clear collapse in reduced units, indicating that the normal stress differences (configurational parts) are at least as isomorph invariant as the strain-rate dependent part of the pressure.} \label{NormalStsDiffs}
\end{figure}

Finally we present data for the normal stress difference $(\sigma_{xx}-\sigma_{yy})/2$, an important quantity in non-linear rheology\cite{thinning}. In complex fluids the normal stress difference can be a probe of microstructure\cite{Mall-Gleissle/others:2002}, while in simulations evaluating the normal stress difference has been used to judge the validity of flow algorithms, for example by Hoover et al\cite{hoover}. These authors discuss the validity of the DOLLS and SLLOD algorithms by comparison to the ``correct'' boundary driven flow for simple shear. They find that neither SLLOD nor DOLLS reproduces the correct normal stress differences -- while SLLOD tends to get the correct sign, their size can be too small by an order of magnitude. We do not wish to enter the discussion of which algorithm is ``correct''; our focus is the isomorph invariance of the SLLOD algorithm. 

By the same reasoning that argued for the approximate isomorph invariance of the strain-rate dependent parts of pressure and energy, we expect the configurational parts of the normal stress differences are isomorph invariant. Data confirming this for SCLJ are shown in Fig.~\ref{NormalStsDiffs}. Note that Ref.~\onlinecite{hoover} noted kinetic contributions to normal stress differences, which in some situations dominate the potential ones. We did not consider the kinetic terms since the isomorph theory says nothing about them (also they are not recorded by our molecular dynamics software).

\section{Discussion}\label{summary}

We have investigated the isomorph theory's predictions for the SCLJ and KABLJ systems undergoing steady shear flow. Both model systems are simple in the Roskilde sense of the term\cite{prx} (i.e., strongly correlating) liquids and known to have good isomorphs at zero strain rate referring to the standard two-dimensional thermodynamic phase diagram. This paper has demonstrated that the isomorph concept extends to steady-state non-equilibrium situations described by the SLLOD equations of motion, for which the phase diagram is three dimensional because the strain rate defines an extra dimension of the phase diagram.

We studied structure, dynamics, and rheology in steady-state Couette shear flows. As expected, the structures of both systems were unaffected by shear at low strain rates, but a change of structure was observed at the onset of nonlinear effects. The range of strain rates considered was large enough to capture genuine shear-thinning behavior. It is significant that our results include this nonlinear regime, since the isomorph invariance of transport coefficients in the linear regime follows from that of the equilibrium properties. We obtained simulation results for structure studied via the pair-correlation function, dynamics studied via the incoherent intermediate scattering function, and transport quantities studied via the steady-state viscosity and the normal stress difference. The results show that the proposed extension of the equilibrium isomorph theory describes well SLLOD steady-state non-equilibrium situations.

Although potential energy and pressure are not inherently isomorph invariant, the strain-dependent parts of the reduced potential energy and (to a lesser extent) pressure are invariant when considered as functions of reduced shear rate. Data published by Ge {\it et al.} \cite{Ge} are consistent with our results. They showed that for a dense LJ liquid under shear flow, the potential energy and the pressure can be fitted by a power-law dependence on strain rate,

\be 
U=U_0+a\dot{\gamma}^{\alpha} 
\ee
\be 
P=P_0+b\dot{\gamma^{\alpha}} \,,
\ee
in which $\alpha$ is a common exponent that depends on density and temperature. They found\cite{Ge} that the linear expression $\alpha=A+BT-C\rho$ represents well their simulations with $A=3.67$, $B=0.69$, and $C=3.35$. To make a connection between these results and isomorph theory, recall that the collapse seen in Fig. \ref{fig4} and (to a lesser extent) in Fig. \ref{fig5} is a consequence of the isomorph theory and the additional assumption that the term $g(Q)$ does not depend on strain rate (see the discussion of projected isomorphs around those figures). The master curves contain all the information about the strain-rate dependence of these quantities, and so any quantity characterizing such a master curve -- for example a power-law exponent -- is uniquely associated with the projected isomorph, the equilibrium isomorph. Equivalently, the exponent determined by varying strain-rate at different points in the $(\rho, T)$ plane must be invariant along equilibrium isomorphs. Thus the theory implies that $d\alpha=0$ along an (equilibrium) isomorph and, in particular, the strain-rate exponent must be the same for the potential energy and the pressure. This means that one can write

\be 
\Delta \tilde{U}=\frac{U-U_{0}}{k_{B}T}=\tilde{a}(\tilde{\dot{\gamma}})^\alpha 
\ee

\be 
\Delta \tilde{P}=\frac{P-P_{0}}{k_{B}T}=\tilde{b}(\tilde{\dot{\gamma}})^\alpha 
\ee
in which $\tilde{a}=a/(Tt_0^\alpha)$ and $\tilde{b}=b/(Tt_0^\alpha)$. The linear expression of Ge {\it et al.} $\alpha=A+BT-C\rho$ implies that $\alpha$ is constant along straight lines in the $\rho,T$ plane. According to the isomorph theory, however, their data would be even better matched by the almost straight lines in the $(\ln\rho, \ln T)$ plane defining the isomorphs. More simulations are needed to test this prediction, but based on the available data we can already note the following. The isomorph theory implies that $d\alpha=0$ along an isomorph, i.e., $BdT-Cd\rho=0$. Based on this one can estimate the density-scaling exponent from the data of Ge {\it et al.} from the density-temperature variation along an isomorph: $\gamma=d\ln T /d\ln\rho=(\rho/T)(dT/d\rho)\simeq (0.8/1) C/B\simeq 4$ which given the uncertainties is consistent with our findings.

The fact that isomorph invariance extends beyond equilibrium situations could provide a powerful tool to check theories of non-equilibrium behavior. This is because isomorph invariance imposes a constraint on the temperature and density dependence of transport coefficients, and any general theory for these must result in an isomorph invariant expression for the reduced transport coefficients. This would be analogous to the ``isomorph filter'' for theories of the dynamics of viscous liquids approaching the glass transition \cite{IV}.

\acknowledgments

The authors are indebted to Trond Ingebrigtsen for several helpful comments. The center for viscous liquid dynamics ``Glass and Time'' is sponsored by the Danish National Research Foundation's grant DNRF61.

\bibliography{SLLOD}

\end{document}